\documentclass[preprint, 5p]{elsarticle}

\pdfpageattr{/Group << /S /Transparency /I true /CS /DeviceRGB >>}

\usepackage[utf8]{inputenc}
\usepackage[T1]{fontenc}
\usepackage{textcomp}
\usepackage{fixltx2e}
\usepackage{microtype}
\usepackage{calc}
\usepackage[normalem]{ulem}

\usepackage{xcolor} 
\usepackage{footnote}
\usepackage{csquotes}

\DeclareUnicodeCharacter{2212}{-} 

\usepackage{mathtools}
\usepackage{amsmath,amsfonts}
\usepackage{algorithmic}

\usepackage{tikz}
\usetikzlibrary{spy}

\usepackage{booktabs}
\usepackage{url}
\usepackage{multirow}

\usepackage[hidelinks]{hyperref}

\usepackage{lipsum}
\usepackage{soul}

\usepackage{csquotes}
\newcommand{\textcite}[1]{\citet{#1}}

\usepackage{amsmath}

\usepackage{amssymb}
\usepackage{amsfonts}
\usepackage{amstext}
\usepackage{amsthm}

\usepackage{booktabs}
\usepackage{multirow}
\usepackage{url}

\usepackage[inline]{enumitem}

\usepackage[caption=false,font=footnotesize]{subfig}
\usepackage{graphicx}
\usepackage{pgfplots}
\DeclareGraphicsExtensions{.pdf,.png,.jpg}
\DeclareGraphicsExtensions{.pdf,.png,.jpg,.tikz}
\usepackage{tikz}
\usetikzlibrary{arrows}
\usetikzlibrary{calc}
\usetikzlibrary{chains}
\usetikzlibrary{scopes}

\usepackage[american]{babel}
\usepackage{hyphenat}

\usepackage[]{algorithmic}
\usepackage{algorithm}

\usepackage[capitalize,noabbrev]{cleveref}
\crefformat{footnote}{#2\footnotemark[#1]#3}
\Crefformat{figure}{#2Fig.~#1#3}
\Crefformat{equation}{#2Eq.~(#1)#3}

\usepackage{todonotes}

\usepackage{comment}

\usepackage[color]{changebar}
\def\todo{%
    \cbcolor{magenta}
    \cbstart%
    \begingroup
    \color{magenta}
    \obeylines%
    \begingroup\lccode`~=`\^^M\lowercase{\endgroup\def~}{\par\leavevmode}%
    \parindent0em%
    \catcode`\_=\active
    \catcode`\<=\active\lccode`~=`<\lowercase{\def~}{$<$}%
    \catcode`\>=\active\lccode`~=`>\lowercase{\def~}{$>$}%
    \catcode`\#=\active\lccode`~=`\#\lowercase{\def~}{$\#$}%
    \catcode`\^=\active\lccode`~=`\^\lowercase{\def~}{$\hat{~}$}%
    \todoCtd
    }\def\todoCtd#1{%
    TODO: #1%
\ifx&#1&...\fi%
    \endgroup
    \cbend
    \relax
}

\NewDocumentCommand\IEEE{ s m d[] }{%
    \IfBooleanTF{#1}{}{IEEE\,}
    \nolinebreak[2]
    #2%
    \IfNoValueTF{#3}{%
        }{%
        \StrGobbleLeft{#3}{1}[\sommerIEEEFirstLetter]%
        \IfEq{\sommerIEEEFirstLetter}{}{%
            #3
            }{%
            \nolinebreak[3]
            \StrLeft{#3}{1}%
            \sommerIEEELettersSlashed{\sommerIEEEFirstLetter}%
        }%
    }%
}
\newcommand{\sommerIEEELettersSlashed}[1]{%
    /
    \StrLeft{#1}{1}%
    \StrGobbleLeft{#1}{1}[\sommerIEEESubsequentLetter]%
    \IfEq{\sommerIEEESubsequentLetter}{}{%
        }{%
        \sommerIEEELettersSlashed{\sommerIEEESubsequentLetter}
    }%
}

\usepackage{todonotes}
\usepackage{import} 

\usepackage[
  range-phrase=--,
  per-mode=symbol-or-fraction,
  binary-units=true,
  range-units=single,
  list-units=single,
  detect-all,
  list-final-separator={, and },
]{siunitx}
\usepackage{silence}
\DeclareSIUnit\decibelm{dBm}

\RequirePackage{xstring}
\RequirePackage{xparse}
\RequirePackage[]{acro}
\NewDocumentCommand\acrodef{mO{#1}mG{}}{\DeclareAcronym{#1}{short={#2}, long={#3}, #4}}
\acrodef{5G}{fifth generation}
\acrodef{ACC}{adaptive cruise control}
\acrodef{AIFS}{arbitration inter-frame space}
\acrodef{AoI}{age of information}
\acrodef{BER}{bit error rate}
\acrodef{BPSK}{binary phase shift keying}
\acrodef{BRR}{beacon reception ratio}
\acrodef{BTR}{beacon transmission ratio}
\acrodef{CACC}{cooperative adaptive cruise control}
\acrodef{CAM}{cooperative awareness message}
\acrodef{CBR}{channel busy ratio}
\acrodef{CC}{cruise control}{alt={cruise controller}}
\acrodef{CSMA/CA}{carrier sense multiple access with collision avoidance}
\acrodef{C-V2X}{cellular V2X}
\acrodef{CW}{contention window}
\acrodef{DSRC}{distributed short-range communication}
\acrodef{eCDF}{empirical cumulative distribution function}
\acrodef{GPS}{global positioning system}
\acrodef{ICA}{intersection collision avoidance}
\acrodef{IDM}{intelligent driver model}
\acrodef{INR}{informed neighbors ratio}
\acrodef{ITS}{intelligent transportation system}{short-plural=}
\acrodef{IVC}{inter-vehicle communication}
\acrodef{KNR}{known neighbors ratio}
\acrodef{LOS}{line-of-sight}
\acrodef{MAC}{medium access control}
\acrodef{NLOS}{non line-of-sight}
\acrodef{PAoI}{peak age of information}
\acrodef{PID}{proportional-integral-derivative}
\acrodef{PLCP}{physical layer convergence protocol}
\acrodef{PRR}{packet reception ratio}
\acrodef{RSU}{roadside unit}{short-indefinite={an}}
\acrodef{SNIR}{signal-to-noise-and-interference-ratio}
\acrodef{SNR}{signal-to-noise ratio}
\acrodef{V2X}{vehicle-to-everything}
\acrodef{SDR}{software defined radio}

\newcommand{\p}{\IEEE{802.11}[p]}

\newcommand{\theoryAlpha}{0.75}
\newcommand{\theoryBeta}{0.005}

\AtBeginDocument{%
  \providecommand\BibTeX{{%
    \normalfont B\kern-0.5em{\scshape i\kern-0.25em b}\kern-0.8em\TeX}}}

\begin{document}

\begin{frontmatter}

\title{%
Focusing on Information Context for ITS using a Spatial Age~of~Information Model
}

\author{Julian Heinovski\corref{cor1}}
\ead{heinovski@ccs-labs.org}
\author{Jorge Torres G\'omez}
\ead{torres-gomez@ccs-labs.org}
\author{Falko Dressler}
\ead{dressler@ccs-labs.org}
\cortext[cor1]{Corresponding author}
\address{%
School of Electrical Engineering and Computer Science, TU Berlin, Germany
}


\begin{abstract}
New technologies for sensing and communication act as enablers for cooperative driving applications.
Sensors are able to detect objects in the surrounding environment and information such as their current location is exchanged among vehicles.
In order to cope with the vehicles' mobility, such information is required to be as fresh as possible for proper operation of cooperative driving applications.
The \ac{AoI} has been proposed as a metric for evaluating freshness of information; recently also within the context of \acp{ITS}.
We investigate mechanisms to reduce the \ac{AoI} of data transported in form of beacon messages while controlling their emission rate.
We aim to balance packet collision probability and beacon frequency using the average \ac{PAoI} as a metric.
This metric, however, only accounts for the generation time of the data but not for application-specific aspects, such as the location of the transmitting vehicle.
We thus propose a new way of interpreting the \ac{AoI} by considering information context, thereby incorporating vehicles' locations.
As an example, we characterize such importance using the orientation and the distance of the involved vehicles.
In particular, we introduce a weighting coefficient used in combination with the \ac{PAoI} to evaluate the information freshness, thus emphasizing on information from more important neighbors.
We further design the beaconing approach in a way to meet a given \ac{AoI} requirement, thus, saving resources on the wireless channel while keeping the \ac{AoI} minimal.
We illustrate the effectiveness of our approach in Manhattan-like urban scenarios, reaching pre-specified targets for the \ac{AoI} of beacon messages.
\end{abstract}

\begin{keyword}
Age of information; vehicular networking; cooperative driving; intelligent transportation systems; beaconing
\end{keyword}

\end{frontmatter}

%

\acresetall
\section{Introduction}%
\label{sec:introduction}

Vehicles are becoming more capable in terms of sensors and computing power, we can already see first deployments of \acp{ITS}.
They are able to do real-time monitoring of their surrounding environment, which allows detection of objects such as other vehicles or pedestrians.
Such a detection can help in taking driving decisions and might be performed collaboratively among vehicles.
\Ac{V2X} communication technologies such as \p{} (often referred to as \ac{DSRC}), and \ac{C-V2X} allow exchanging of information with other vehicles or roadside infrastructure.
Such communication capability enables cooperative driving applications, which bring a set of new features and services for today's driving, but also demands new communication strategies for proper operation \cite{dressler2019cooperative}.

Such applications require fresh (up-do-date) information from other vehicles to operate properly.
Due to the inherent mobility of \acp{ITS}, location-information may become outdated and eventually no longer relevant to an application because of position changes.
To ensure up-to-date information, vehicles exchange their location with regular \emph{beacons} messages, such as \acp{CAM}~\cite{etsi_302637-2-v132}.
Freshness of such information does not only depend on frequent beacons' transmissions but also on the network capacities, which can lead to packet collisions and the loss of data.

Balancing requirements to update beacons regularly and network capacities, the concept of the \acf{AoI} provides a framework to evaluate information freshness \cite{yates2021age,kosta2017age}.
The \ac{AoI} metric complements raw delay, loss, and throughput in the network, which in turn accounts for the process of emission and delays introduced in the communication chain, all-together~\cite{yates2021age}.
Inherently, this distinguishes \ac{AoI} metrics from conventional delay metrics \cite{lin2020average}, allowing to optimize of the network freshness as the best balance between throughput and delay.

\begin{figure}
    \centering
    \includegraphics[width=0.68\columnwidth]{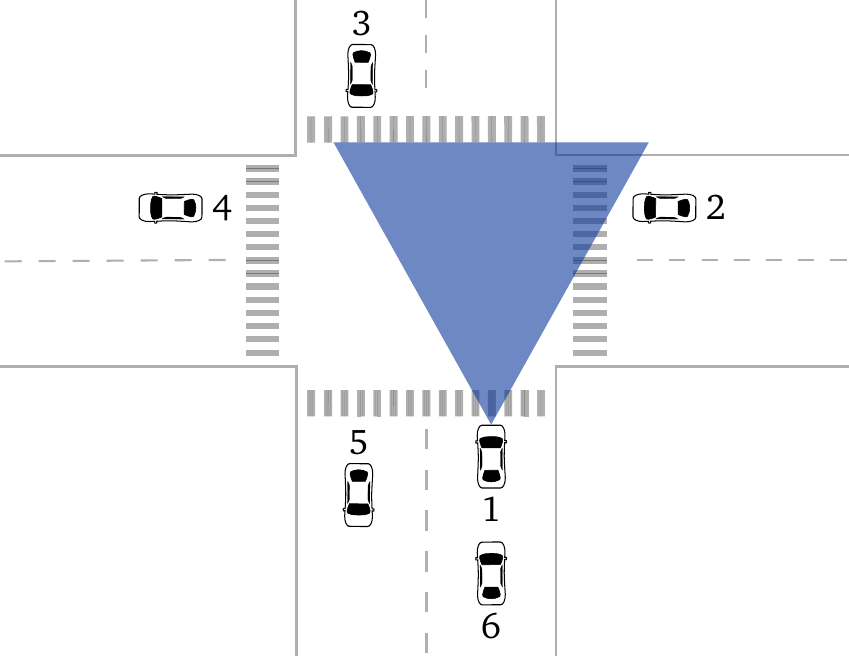}
    \vspace{-.5em}
    \caption{%
        Example scenario:
        Vehicles in the direction of movement of vehicle $1$ (vehicles \numlist{2;3}) are more relevant than others (vehicles \numlist{4;5;6}) when considering a safety application such as \ac{ICA}.
    }%
    \label{fig:scenario_intersection}
\end{figure}

Recent research illustrates the use of \ac{AoI} metrics in the vehicular context.
The average \ac{AoI} and the \ac{PAoI} are used to find the best strategy for the emission rate of beacon packets, see for instance \cite{kaul2011minimizing,librino2016multihop,baiocchi2017model,lyamin2020age,baiocchi2021age2,baiocchi2021age,cao2022optimize} (further discussion in the next Section).
However, these metrics do not consider information context \cite{yates2019age}; not all packets necessarily carry the same information's importance, thereby not introducing the same level of freshness for the status update.
To illustrate the significance of information context, let us consider an \ac{ICA} scenario as depicted in \cref{fig:scenario_intersection}.
Focusing on vehicle \num{1}, the status updates of the vehicles in front (i.e., vehicles \numlist{2;3}) are more critical to avoid potential collisions than those from the other vehicles (i.e., vehicles \numlist{4;5;6}).
Thus, it is preferable to collect updated information about vehicles \numlist{2;3} rather than \numlist{4;5;6} in order for vehicle \num{1} to avoid collisions at this intersection.
In this sense, we intent for an \ac{AoI}-based metric to account for this application-dependent information context.

In the literature, some reported studies address the information's content with the \ac{AoI}.
Examples include techniques
to best predict \textit{Markovian} sources \cite{kam2018towards},
to better synchronize cache content with the remote source \cite{zhong2018two},
to account for only newly arrived information \cite{lin2020average,maatouk2020age},
and to use information-theoretic approaches for reduced uncertainty about the source \cite{yates2021age}.


In a different approach, we consider the transmitting vehicle's location as a metric to quantify the importance of their information.
We emphasize such importance with weighting coefficients to evaluate the average \ac{PAoI} metric in the context of cooperative driving.
These coefficients inherently devise some filtering of received packets, which allows for treating vehicles' information more selectively.
Using our model allows focusing on timely updates of relevant vehicles only for meeting a given \ac{AoI} requirement.

In this paper, which is an extension of the previous work in \cite{heinovski2022spatial}, we further report on the impact of the scenario on the \ac{PAoI} when applying our spatial model.
In addition, we demonstrate the practical use of the proposed model by applying it to an \ac{AoI}-based beacon adaption algorithm.

Our main contributions can be summarized as follows:
\begin{itemize}
    \item We propose a spatial model for interpreting the \ac{AoI} of received packets based on the spatial location of the transmitting vehicle, making \ac{AoI} context specific.
    \item We evaluate the \ac{AoI} and the impact of our model on individual communication links and the overall network.
    \item We show that using our model helps controlling the beacon rate necessary for achieving a given \ac{AoI} requirement.
    \item We demonstrate the practical use of the proposed model by applying it to an \ac{AoI}-based beacon adaption algorithm.
\end{itemize}

Elaborating on the above contribution, the paper is structured as follows.
We discuss in further detail the use of \ac{AoI} metrics in the vehicular context in \cref{sec:aoi}.
We present the new \ac{AoI} model and introduce the weighting coefficient in \cref{sec:spatial_model} with the corresponding analytic description.
We also introduce in \cref{sec:adaptive-beaconing} an adaptive mechanism for the beacon rate with the proposed \ac{AoI} model using a \ac{PID} controller.
In \cref{sec:evaluation}, we assess the performance of the proposed model and the adaptive algorithm to meet freshness requirements.
Finally, we sketch some concluding remarks in \cref{sec:conclusion}.

%

\section{Age of Information in ITS}%
\label{sec:aoi}

Following the standard \p{}, we already see a number of studies that address the use of \ac{AoI} metrics for time-critical applications in vehicular networks \cite{kaul2011minimizing,librino2016multihop,baiocchi2017model,lyamin2020age,baiocchi2021age2,baiocchi2021age,cao2022optimize} and including test-bed demonstrators \cite{jimenezsoria2022experimental}.
The \ac{AoI} metric is reported to update the network freshness for the exchange of vehicles' speeds and positions.
Some of these works provide closed-form expressions for the \ac{AoI} metric \cite{baiocchi2017model,lyamin2020age,baiocchi2021age2,baiocchi2021age}, while other works estimate the average \ac{AoI} metric numerically \cite{kaul2011minimizing,librino2016multihop}.




Accounting for the \ac{AoI} metrics, other works use analytic methods for the average \ac{AoI} \cite{lyamin2020age} and \ac{PAoI} \cite{cao2022optimize}.
The average \ac{AoI} metric is formulated mainly in two different approaches.
On one hand, \citet{lyamin2020age,vinel2009estimation} straightly formulate the average \ac{AoI} as the average of the time duration between two consecutive received packets.
They assume that the time duration distributes according to the joint event where two transmissions do not collide in the channel.
The channel collision probability is evaluated according to the formulation provided by \citet{vinel2009estimation}.

On the other hand, the average \ac{AoI} is evaluated using the formula for the remaining service time in a queue \cite{baiocchi2017model,baiocchi2021age} and considering the hidden \cite{baiocchi2017model,baiocchi2021age} and non-hidden node problem scenario \cite{baiocchi2021age2}.
In the hidden node scenario, the time duration of message transmissions is expanded, assuming that hidden nodes transmit independently with a random phase between \num{0} and the transmission duration parameter.
In the non-hidden scenario, \citet{baiocchi2021age2} also derive a formula for node and network levels.
The node-level accounts for the average \ac{AoI} at any arbitrary node, assuming they only transmit the most recent packet.
The network level evaluates the case where nodes do not transmit new packets till the current one is sent.
In this case the network is modeled according to a Markov chain model, the transition probabilities can be derived~\cite{vinel2009estimation}.


In a different direction, based on simulation results, the existence of a unique beacon period minimizing the average \ac{AoI} is illustrated for a certain number of vehicles, and \ac{CW} sizes \cite{kaul2011minimizing}.
Following these results, a rate control algorithm is derived from adapting the broadcast period based on local measurements of the average \ac{AoI}.
The vehicle reduces or increases the beacon period by comparing it to the estimated average \ac{AoI} metric looking for the maximum network freshness.

All the above studies conduct simulations based on the \p{} standard for single-hop \cite{kaul2011minimizing,baiocchi2017model,baiocchi2021age,lyamin2020age,baiocchi2017model} and multi-hop~\cite{librino2016multihop,baiocchi2017model,kaul2011minimizing,librino2016multihop} networks.
Besides, a variety of scenarios for the traffic of vehicles have been studied.
Examples include four lane roads~\cite{kaul2011minimizing}, highways \cite{cao2022optimize,jimenezsoria2022experimental}, platooning \cite{librino2016multihop,lyamin2020age}, the more artificially Manhattan Grid~\cite{baiocchi2017model,baiocchi2021age,jimenezsoria2022experimental}, the TapasCologne scenario~\cite{baiocchi2017model,baiocchi2021age}, and open environments \cite{baiocchi2021age2}.
Furthermore, the work in \cite{jimenezsoria2022experimental} recreates a traffic scenario integrating simulators with \ac{SDR} devices.

However, these reported solutions are only addressing protocol parameters (e.g., beacon rate, \ac{CW} size) and the channel impact (e.g., collisions, noise) to formulate the average \ac{AoI} metric.
Addressing information context, \citet{michalopoulou2021framework} seek to minimize the information aging in the spatial dimension when evaluating the product of the speed of the vehicle and the time duration between received packets.
The information age is reduced by stating the optimization problem in the spatial domain to minimize the predicted location error.
Also in the spatial domain, \textcite{parella2022adaptive} implement a penalty function for the \ac{AoI}, addressing a better prediction trajectory of neighboring vehicles.
The penalty function (dependent on the \ac{AoI} of \ac{CAM} messages) is defined using the distance difference between the true and predicted position of the neighboring vehicles.
The importance of information can also be considered as a weight factor on the average \ac{AoI} metric.
\textcite{zhang2022aoioriented} evaluate the weight while balancing the message's priority, persistence, and reliability.

In a different approach, we introduce a degree of importance in the \ac{AoI} metrics concerning the intended direction of the vehicle and its surrounding.
In the form of weighting coefficients as formulated by \citet{sorkhoh2020minimizing}, we incorporate into the average \ac{PAoI} metric the vehicular context (direction, surrounding), looking for some meaning of received beacon packets \cite{kosta2017age}.
In doing so, we study the \ac{PAoI} metric weighting as more critical for those vehicles in the direction of movement (cf.\ \cref{fig:scenario_intersection}) and with less importance otherwise, when considering a safety application such as \ac{ICA}~\cite{chen2016survey} as an example use case.
We use this modified \ac{PAoI} metric and compare it to a similarly modified \ac{AoI} requirement of \SI{100}{\ms}, which is often used in literature as desired update interval \cite{ploeg2011design,heinovski2019modeling}.
We thus identify how many vehicles have fresh information, similar to using the original \ac{AoI} definition.

%

\section{A Spatial Model for the AoI}%
\label{sec:spatial_model}

%
%

Typically, the \ac{AoI} metrics are measured per user irrespective of their location.
All the packets received from surrounding vehicles are treated with equal importance.
However, the level of importance is application dependent:
E.g., while in platooning only the members of the platoon itself are relevant, it is the surrounding vehicles within the direction of movement that are important for safety-related applications such as \ac{ICA}.
As one example use case, \cref{fig:scenario_intersection} thus shows such an intersection scenario.
Here, the most valuable information for vehicle \num{1} will be located in the direction of its movement (shadowed area).
Thus, vehicles within this area (i.e., vehicles \numlist{2;3}) should be assigned a higher level of importance than other vehicles in the surrounding.
Beacon packets coming from vehicles in the rear of the intended direction will not be that informative about the traffic in the intended direction of vehicle \num{1}.
Therefore, vehicles in front will be more demanded to reduce the corresponding \ac{AoI} metrics than the vehicles in the rear.
Since all vehicles are equal in the standard \ac{AoI}, frequent beacon transmissions from the less important vehicles can lead to unnecessary channel load in this case.
If the communication protocol was aware of this application-specific level of importance, it could update the periodicity of the beacons in accordance and eventually reduce the load on the wireless channel.

\subsection{Weighted Peak AoI}%
\label{sec:coefficient_calculation}



\begin{figure}
\centering
\includegraphics[width=\columnwidth]{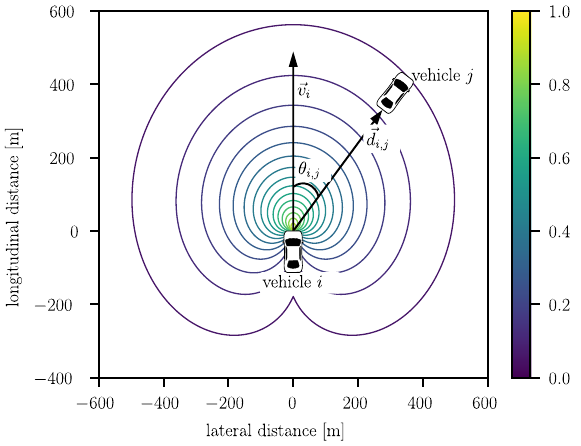}
\vspace{-1.5em}
\caption{%
    Visual representation of our spatial model for calculating the weighting coefficient used for the \ac{AoI} interpretation with an example configuration of $\alpha = \theoryAlpha{}, \beta = \theoryBeta{}$.
}%
\label{fig:spatial_model}
\end{figure}

To consider the location of the transmitting vehicle as a metric of importance to the information, we propose a new way of interpreting the \ac{AoI}.
To that end, we introduce a weighting coefficient that is applied to the \ac{PAoI} metric as well as to an \ac{AoI} requirement, emphasizing on packets from important vehicles.
Thereby, we introduce some level of selectivity for the received packets which allows to treat vehicles' information differently according to the importance to the application.
Our model can be easily adjusted to the requirement of a specific application through parameters and does not modify the underlying \ac{AoI} metric itself.

We choose the weighting coefficient as a raised-cosine function with a decay factor as
\begin{align}\label{eq:weighting_coefficient_omega}
    \omega_{j,i}=
\frac{1}{2}\left(1+\cos(\alpha\theta_{i,j})\right)e^{-\beta \parallel \vec{d}_{i,j}\parallel } \quad ,
\end{align}
where $\theta_{i,j}$ is the angle between the transmitting vehicle $j$ and the direction of movement of vehicle $i$ and $\parallel \vec{d}_{i,j} \parallel$ is the distance between both vehicles, while $\alpha$ and $\beta$ are two coefficients to select the degree of selectivity in the spatial domain.
The coefficient $\alpha$ provides selectivity in the radial direction, while $\beta$ in the azimuth direction.
The larger the value of $\alpha$ or $\beta$ is, the narrower is the beam of vehicle~$i$.
\Cref{fig:spatial_model} shows a visual representation of our weighting coefficient with an example configuration of $\alpha = \theoryAlpha{}, \beta = \theoryBeta{}$.

To derive the angle and the distance between vehicles, we assume that vehicles are equipped with \ac{GPS} receivers, and that this information is exchanged between vehicles in periodic beaconing messages.
Using the configuration in \cref{fig:spatial_model}, $\omega_{j,i}$ is close to \num{1} whenever vehicle $j$ is in the direction of movement of vehicle $i$.
Otherwise, it is close to \num{0} whenever vehicle $j$ is away, being \num{0} when vehicle $j$ is located at the rear of vehicle $i$.

\begin{figure}
\centering
\includegraphics[width=\columnwidth]{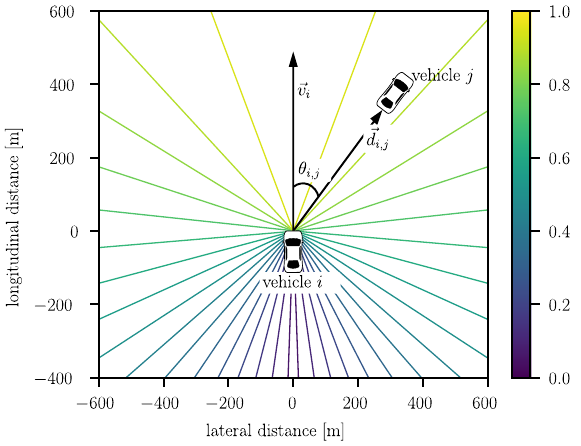}
\vspace{-1.5em}
\caption{%
    Visual representation of our spatial model for calculating the weighting coefficient used for the \ac{AoI} interpretation with an example configuration of $\alpha = 0.5, \beta = 0$.
    This configuration only takes the angle between vehicles $i$ and $j$ into account.
}%
\label{fig:spatial_model_angle_only}
\end{figure}

\begin{figure}
\centering
\includegraphics[width=\columnwidth]{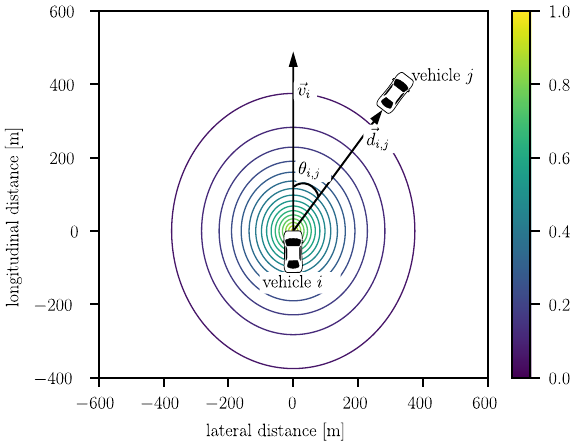}
\vspace{-1.5em}
\caption{%
    Visual representation of our spatial model for calculating the weighting coefficient used for the \ac{AoI} interpretation with an example configuration of $\alpha = 0, \beta = 0.0075$.
    This configuration only takes the distance between vehicles $i$ and $j$ into account.
}%
\label{fig:spatial_model_distance_only}
\end{figure}

In order to parameterize our proposed model, one needs to select values for the model parameters $\alpha$ and $\beta$.
Since these will determine the relevance of vehicles' information, their values have to be selected carefully and application-dependent.
In order to give further intuition on the behavior of these, \cref{fig:spatial_model_angle_only,fig:spatial_model_distance_only} show visual representations of two more example configurations.
In \cref{fig:spatial_model_angle_only}, the distance between vehicles is not used for calculating the weighting coefficient $\omega_{j,i}$ due to a value of \num{0} for $\beta$.
Thus, the area around the vehicle $i$ is mapped to the coefficient $\omega_{j,i}$ by the cosine-part of \cref{eq:weighting_coefficient_omega}, using the angle towards vehicle $j$.
Here, a close vehicle and a far-away vehicle will receive the same value for $\omega_{j,i}$ if their angle towards vehicle $i$ is the same.
In contrast, in \cref{fig:spatial_model_distance_only}, the angle between vehicles is not used for calculating the weighting coefficient due to a value of \num{0} for $\alpha$.
Thus, the distance from vehicle $i$ to other vehicles is mapped to the coefficient $\omega_{j,i}$ by part with Euler's number of \cref{eq:weighting_coefficient_omega}.

With this coefficient, we measure the importance of the introduced age per received packet using the average \ac{PAoI} metric as
\begin{equation}\label{eq:p_aoi_link}
    \Delta^{(\omega)}_{j,i}=\omega_{j,i}\Delta^{(p)}_{j,i} \quad ,
\end{equation}
where $\Delta^{(p)}_{j,i}$ denotes the average \ac{PAoI} metric for a link between vehicles $j$ and $i$.
Correspondingly, the combined \ac{PAoI} per neighboring vehicle $j$ at vehicle $i$ will be calculated after averaging the perceived $\Delta^{(p)}_{j,i}$ as
\begin{equation}\label{eq:p_aoi_neighbors}
    \Delta^{(\omega)}_{i}=\frac{1}{N-1}\sum_{j=1}^{N-1}{\omega_{j,i}\Delta^{(p)}_{j,i}} \quad .
\end{equation}
Finally, we account for the network operation after averaging for the total of nodes as
\begin{equation}\label{eq:p_aoi_network}
    \Delta^{(\omega)}=\frac{1}{N(N-1)}\sum_{i=1}^{N}\sum_{j=1}^{N-1}{\omega_{j,i}}\Delta^{(p)}_{j,i} \quad .
\end{equation}

The average \ac{PAoI} ($\Delta^{(p)}_{j,i}$) can be derived by analytical or numerical means through simulators.
Analytically, it can be obtained after calculating the expected average of the inter-arrival ($Y_n$) and system time ($T_n$) as \mbox{$\Delta^{(p)}_{j,i}=\mathbb{E}\{Y\}+\mathbb{E}\{T\}$} when all packets emitted are received \cite{yates2019age}.
However, some packets will not be successfully received due to the system and channel conditions (e.g., collisions, replacement of the old beacon frames, low reliability during their reception, etc.) \cite{vinel2009estimation}.
Taking into account the impact of the system and channel effects in the packet reception process, as given by the successful probability $P_{sd}$, the average \ac{PAoI} can be calculated as
\begin{equation}\label{eq:p_aoi}
    \Delta^{(p)}_{j,i}=\frac{1}{P_{sd}}\mathbb{E}\{Y\}+\mathbb{E}\{T\} \quad ,
\end{equation}
where the value for $P_{sd}$ in \cref{eq:p_aoi} can be directly computed considering the impact of noise and collisions using the closed-form expressions in \cite[Eq. (7)]{vinel2009estimation}, or through simulations.

The \cref{eq:p_aoi} can be directly computed by recalling the use of the negative binomial distribution as described in~\cite{torres-gomez2022age}.
However, it can be intuitively derived based on the meaning of the related variables in \Cref{eq:p_aoi}.
$P_{sd}$ can be interpreted as the ratio of successfully transmitted packets; thus, its inverse will provide the total of attempts to have a successful transmission.
Therefore, the first term in \cref{eq:p_aoi} will provide the waiting period before a packet is successfully transmitted.
Adding the average time spent on the system ($\mathbb{E}\{T\}$) will thus provide the average \ac{PAoI}.

\subsection{Remarks}%
\label{sec:model_remarks}

The introduced coefficients in \cref{eq:p_aoi_neighbors} provide a mean to ``filter'' packets according to their relevance.
For instance, in the intersection scenario depicted in \cref{fig:scenario_intersection}, the average \ac{PAoI} of packets from the vehicles \numlist{4;5;6} will be lowered as less relevant, thus emphasizing those packets from vehicles at the front side of vehicle \num{1} (vehicles \numlist{2;3}).
In this way, the resulting average \ac{PAoI} will be characterized the most by those links of interest according to the application context.

\Cref{eq:p_aoi_neighbors} is also useful in different \ac{ITS} scenarios, and, also with a different dependency for the coefficients other than \cref{eq:weighting_coefficient_omega}.
Overall, the \cref{eq:p_aoi_neighbors} comprises a mean to emphasize some communication links in contrast to others.
Once the links of interest are determined, they will shape the resulting average \ac{PAoI} whenever their corresponding coefficients are close to \num{1}.
In contrast, those links whose coefficients are close to \num{0} will not contribute to the \acl{AoI} metric.

Besides, we selected the dependency of the coefficients with the spatial coordinates in \cref{eq:weighting_coefficient_omega} as a two-dimensional function in two separable terms.
One dimension for the azimuth direction defines the raised-cosine function \cite{carlson2002communication}, which conveniently allows multiplying by \num{0} to those packets coming from the rear side of vehicle $i$.
The second dimension is in the radial direction and defines a decay factor, which decrements as long as the distance increases.
Overall, both terms let to a function that is also all-orders differentiable, which accounts for its mathematical tractability.





\subsection{Weighted Target AoI}%
\label{sec:model_aoi_requirement}


The derived weighted \ac{PAoI} metric can be used to fairly evaluate the freshness of the status updates with a given target, i.e., when the age of received packets is less than a given threshold.
This approach is particularly relevant when we want to save resources looking at the \ac{PAoI} metric just performing below a given threshold $T$ (target).
In this way, we avoid the network to operate on the minimum average where demanding resources are higher.
After applying the same weighting coefficient (cf.\ \cref{eq:weighting_coefficient_omega}) to a given threshold $T$ by
\begin{equation}\label{eq:target_link}
    T^{(\omega)}_{j,i} = {\omega_{j,i}}T \quad ,
\end{equation}
we can compare the derived average \ac{PAoI} with the weighted target $T^{(\omega)}_{j,i}$
\begin{equation}\label{eq:target_link_leq}
    \Delta^{(\omega)}_{j,i} \leq T^{(\omega)}_{j,i}
\end{equation}
for a link between vehicles $j$ and $i$.
A single vehicle $i$ can to this for all of its neighboring vehicles $j$ by
\begin{equation}\label{eq:target_neighbors}
    T^{(\omega)}_{i} = \frac{1}{N-1} \sum_{j=1}^{N-1}{\omega_{j,i}}T_{j,i}
\end{equation}
and
\begin{equation}\label{eq:target_neighbors_leq}
    \Delta^{(\omega)}_{i} \leq T^{(\omega)}_{i} \quad .
\end{equation}
Correspondingly, we account for the network operation after averaging for the total of nodes as
\begin{equation}\label{eq:target_network}
    T^{(\omega)} = \frac{1}{N(N-1)}\sum_{i=1}^{N}\sum_{j=1}^{N-1}{\omega_{j,i}}T_{j,i}
\end{equation}
and
\begin{equation}\label{eq:target_network_leq}
    \Delta^{(\omega)} \leq T^{(\omega)} \quad .
\end{equation}
Further discussions on the utility of these expressions and their interpretation are given within \cref{sec:evaluation}.

%

\section{Adaptive Beaconing based on \acs*{AoI}}%
\label{sec:adaptive-beaconing}

In this section, we aim to show the practical use of our spatial model for cooperative driving by applying it to an algorithm for adapting the beacon rate based on the \acl{AoI}.
We use a simple \acf{PID} controller-based algorithm that controls a vehicle's beacon rate based on \ac{PAoI} measurements from neighboring vehicles.
In contrast to other work from the literature \cite{kaul2011minimizing,ni2018vehicular}, our goal is to keep peak \ac{AoI} values below a fixed requirement (\SI{100}{\ms}), rather than achieving minimal \ac{AoI} for all nodes.
Applying our spatial model should help to focus on relevant vehicles while selecting the beacon rates.

%

\subsection{Communication Setup}
\label{sec:algorithm_input}


\begin{figure}
\centering
\includegraphics[width=\columnwidth]{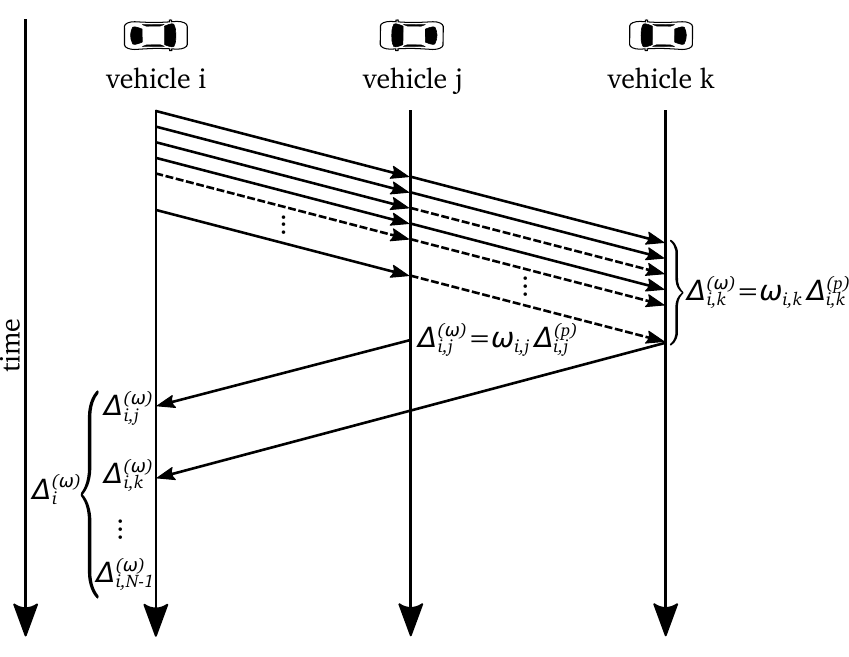}
\vspace{-1.5em}
\caption{Two-hop mechanism of our adaption approach for evaluating the weighted \ac{PAoI} at vehicle $i$.}%
\label{fig:two_hops}
\end{figure}

\Cref{fig:two_hops} shows the two-hop mechanism of our adaption approach which is used to evaluate the weighted \ac{PAoI} $\Delta^{(p)}_i$ locally at vehicle $i$, according to \cref{eq:p_aoi_neighbors}.
Since the information in the beacons of vehicle $i$ are relevant to vehicles $j$ and $k$, the \ac{AoI} as well as the information's relevance (given by the weighting coefficient) needs to be computed at vehicles $j$ and $k$.
As only vehicle $i$ can control its beacon rate, we need this two-hop mechanism to inform $i$ about any required changes.
Note that the indices $i$ and $j$ here are different in comparison to \cref{sec:spatial_model}.

In the first hop, vehicles $j$ and $k$ compute the weighted \ac{PAoI} $\Delta^{(\omega)}_{i,j}$ and $\Delta^{(\omega)}_{i,k}$ upon reception of a beacon from vehicle $i$.
This is done based on \cref{eq:weighting_coefficient_omega} and the spatial locations of the corresponding vehicles.
In the second hop, vehicle $j$ and $k$ transmit these weighted \ac{PAoI} values to vehicle $i$ within their own regular beacons.
In this way, vehicle $i$ is able to compute $\Delta^{(\omega)}_i$ by averaging all weighted \ac{PAoI} measurements from its neighbors $j$ and $k$, following \cref{eq:p_aoi_neighbors}.
$\Delta^{(\omega)}_i$ here corresponds to the freshness of $i$'s information as a combined perspective from vehicle $j$ and $k$.
Following the same procedure, vehicle $i$ also computes the weighted target \ac{AoI} $T^{\omega}_i$ using \cref{eq:target_neighbors}.
Vehicle $i$ is now able to evaluate the freshness of its own information $\Delta^{(\omega)}_i$ perceived at vehicles $j$ and $k$ against the shared requirement $T^{\omega}_i$.
Thus, it can determine whether the information is fresh enough and whether it needs to adjust its beacon rate.


%

\subsection{Beacon Adaption Algorithm}%
\label{sec:pid_controller}

We use a \ac{PID} controller~\cite{minorsky1922directional} on all vehicles to adjust their beacon rate such that \ac{PAoI} values converge to the target \ac{AoI}.
To achieve this, we use the \ac{PAoI} as input and the target \ac{AoI} as set-point for the \ac{PID} controller.
A \ac{PID} controller in general consists of
(1) a \textbf{P}roportional part, which produces a control output proportional to the error between the set-point and the actual output,
(2) an \textbf{I}ntegral part, which accumulates the error over time and generates an output proportional to the integrated error,
and (3) a \textbf{D}erivative part, which produces an output proportional to the rate of change of the error.
The controller output is given by
\begin{equation}\label{eq:pid_controller}
u(t) = G_P \cdot e(t) + G_I \cdot \int e(t)~dt + G_D \cdot \frac{de(t)}{dt} \quad ,
\end{equation}
where
$u(t)$ is the controller output at time $t$,
$e(t)$ is the error (difference to the set-point) at time $t$,
$\int e(t)~dt$ is the integral of the error over time (the accumulated error),
and $\frac{de(t)}{dt}$ is the derivative of the error with respect to time (the rate of change of the error).
$G_P$, $G_I$, and $G_D$ are gain coefficients to control the impact of the individual parts of the \ac{PID} controller towards the output.


As input and set-point for the \ac{PID} controller, we specifically use the average weighted \ac{PAoI} measurement $\Delta^{(\omega)}_i$ and the average weighted target \ac{AoI} $T^{(\omega)}_i$, respectively.
Thereby, we expect the controller to eventually select beacon rates that result in the \ac{PAoI} measurements converging to the target \ac{AoI}.
We implement the individual terms of \cref{eq:pid_controller} in the following way.
The error $e(t)$ at vehicle $i$ in iteration $t$ is given by
\begin{equation}
e(t) = T^{(\omega)}_i - \Delta^{(\omega)}_i
\end{equation}
and the derivate error $de(t)$ evaluated in the discrete domain as
\begin{equation}
\frac{de(t)}{dt} \approx \frac{e(t-\Delta t) - e(t)}{\Delta t} \quad ,
\end{equation}
using only the error from the previous algorithm execution.
%
The gain coefficients for the individual parts of the \ac{PID} controller are chosen by observing the stable system behavior.
We use $G_p = 1.0$, $G_I = 0$, thereby disabling the integral part, and $G_D = 0.1$.

Using \cref{eq:pid_controller}, the adjustment $u(t)$ of the current beacon interval $\lambda_i$ at vehicle $i$ in iteration $t$ is calculated and added to the current beacon interval $\lambda_i$.
We update the new beacon interval within the limits of \SI{10}{\ms} (\SI{100}{\hertz}) and \SI{100}{\s} (\SI{0.01}{\hertz}), such that only valid beacon rates are produced.
The algorithm is executed on a regular basis (i.e., at an interval of $2~\times$ target \ac{AoI}) on all vehicles individually.

%

\section{Evaluation}%
\label{sec:evaluation}

In this section, we evaluate the \ac{PAoI} and the impact of the weighting coefficient $\omega_{j,i}$ from our spatial model (see \cref{sec:spatial_model}).
For this, we perform simulative experiments within the Veins simulator \cite{sommer2011bidirectionally}.
First, we provide an initial analytical assessment of our model in \cref{sec:analytical_results}.
After that, we consider simulative results and discuss the impact of the link distance on the standard \ac{PAoI} in \cref{sec:distance_peak_aoi}.
We continue selecting two specific link distances (i.e., a short and a long one) and analyze the combined \ac{PAoI} without and with our spatial model in \cref{sec:combined_aoi_standard,sec:combined_aoi_spatial}.
Next, we show the impact of our spatial model on the network \ac{PAoI} in \cref{sec:network_aoi}.
Afterwards, we analyze the impact of the model parameters and the scenario on the network \ac{PAoI} in \cref{sec:eval_model_parameters,sec:eval_scenario}.
Finally, we analyze the behavior of our \ac{AoI}-based algorithm for adapting the beacon rate from \cref{sec:adaptive-beaconing} in \cref{sec:adaptive_results}.

\subsection{Initial Analytical Assessment}%
\label{sec:analytical_results}

To provide further intuition on the impact of the coefficient $\omega_{j,i}$, we now perform an initial analytical assessment of the perceived average \ac{PAoI} given by \cref{eq:p_aoi_neighbors}.
We compute it for a given link between \num{200} vehicles that are all moving randomly in a free-space grid.
The corresponding communication parameters are listed in \cref{tab:simulation_parameters_matlab}.
To compute $\Delta^{(p)}_{j,i}$, we use \cref{eq:p_aoi} where the probability of successful beacon reception $P_{sd}$ is obtained from simulation (cf.\ \cref{sec:simulation_setup,sec:distance_peak_aoi}).
We consider a contention-based communication system according to \p{}, where vehicles broadcast beacon messages following a collision avoidance mechanism without retransmissions.
Frames are emitted after verifying free channel access during the \ac{AIFS} and \ac{CW} time windows.
We assume that only the most updated message is queued at the emitter, waiting for a free slot to be transmitted~\cite{baiocchi2017model}.

\begin{table}
    \footnotesize
    \centering
    \begin{tabular}{lr}
        \toprule
        Parameter                                   & Value \\
        \midrule
        Scenario Size                               & \SI{550}{\m} $\times$ \SI{550}{\m} \\
        Simultaneous Vehicles $N$                   & \num{200} \\
        Beacon Size $L$                             & 512 Byte \\
        Bitrate $R$                                 & $\SI{6}{\mega\bit\per\sec}$ \\
        CW size $W$                                 & \numrange{4}{8} \\
        Preamble duration $T_{\text{P}}$            & $\SI{32}{\micro\sec}$ \\
        \acs{PLCP} duration $T_{\text{PLCP}}$       & $\SI{8}{\micro\sec}$ \\
        Propagation delay $\delta$                  & $\SI{1}{\micro\sec}$ \\
        \acs{AIFS}                                  & $\SI{58}{\micro\sec}$ \\
        \bottomrule
    \end{tabular}
    \caption{Parameters used for analytical evaluation}%
    \label{tab:simulation_parameters_matlab}
\end{table}

\begin{table}
    \footnotesize
    \centering
    \begin{tabular}{lr}
        \toprule
        Parameter                                   & Value \\
        \midrule
        Scenario Type                               & Manhattan Grid \\
        Simulation Time                             & \SI{10}{\s} \\
        Beacon Rates                                & \SIlist[list-units=single,list-final-separator = {, }, list-pair-separator= {, }]{1;2;5;8;10;16;20;25;40;50;100}{\hertz} \\
        Carrier Frequency                           & \SI{5.89}{\giga\hertz} \\
        Access Category                             & AC\_VO \\
        EDCA Queue Size                             & 1 \\
        TX Power                                    & \SI{20}{\milli\watt} \\
        Attenuation Model                           & Free-space only ($\alpha = 2$) \\
        \bottomrule
    \end{tabular}
    \caption{Additional parameters used for simulation experiments}%
    \label{tab:simulation_parameters_veins}
\end{table}

\begin{figure}
\centering
\includegraphics[width=\columnwidth]{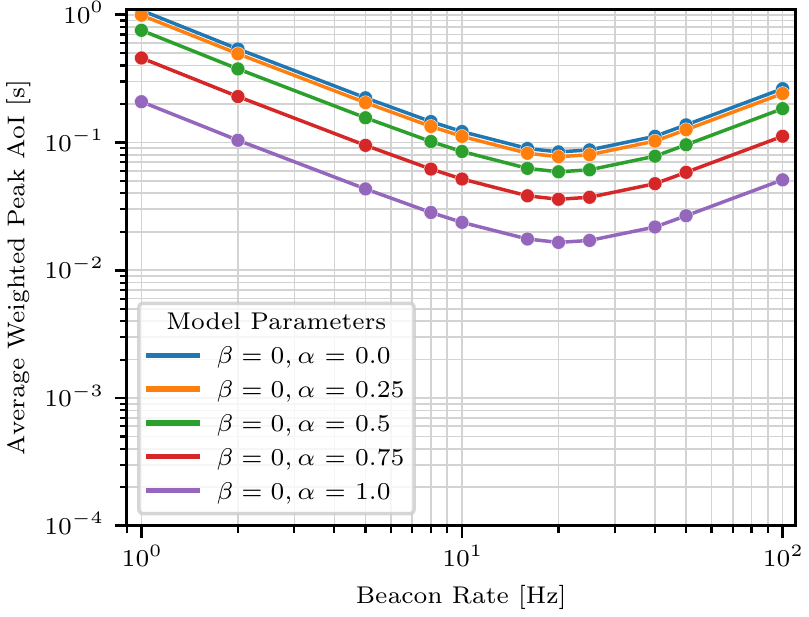}
\vspace{-1.5em}
\caption{%
    Average \ac{PAoI} for different angle coefficients when $\beta=0$.
    The spatial model only focuses on the angle.
}%
\label{fig:AoI_angle}
\end{figure}

\begin{figure}
\centering
\includegraphics[width=\columnwidth]{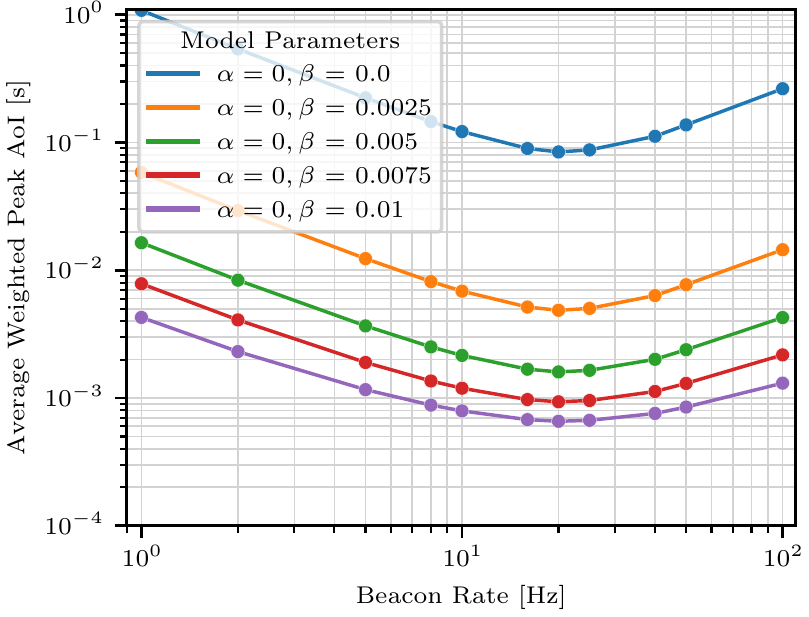}
\vspace{-1.5em}
\caption{%
    Average \ac{PAoI} for different distance coefficients when $\alpha=0$.
    The spatial model only focuses on the distance.
}%
\label{fig:AoI_radius}
\end{figure}

\Cref{fig:AoI_angle,fig:AoI_radius} plot analytical results for the impact of the model parameters $\alpha$ and $\beta$, which define the degree of selectivity for computing the \ac{PAoI} metric.
The case $\alpha=0$, $\beta=0$ results in \mbox{$\omega_{j,i}=1$}, i.e., no spacial selectivity at all, which corresponds to highest \ac{PAoI} metric (standard definition from \cref{eq:p_aoi}).
However, as $\alpha$ and $\beta$ increase, the perceived \ac{PAoI} metrics are reduced due to the reduced importance of those vehicles not in the direction of movement and not that close to the moving vehicle (cf.\ link \num{1}-\num{5} in \cref{fig:scenario_intersection}).
In this case, the contribution of the given link to the total \ac{PAoI} in \cref{eq:p_aoi_neighbors} will become less, thus less critical.
Comparing both figures indicates that the distance has a higher impact on the calculation of the weighting coefficient than the angle.

\subsection{Simulation Setup}%
\label{sec:simulation_setup}

After the initial analytical assessment of our model, we now move on and consider simulative results.
For our simulation, we use the well-known vehicular network simulation framework Veins~\cite{sommer2011bidirectionally} to enable a realistic evaluation.
In particular, we use
OMNeT++ 5.6.2,
SUMO 1.6,
and Veins 5.1.
\Cref{tab:simulation_parameters_matlab,tab:simulation_parameters_veins} together summarize the most important parameters used in our simulations.

\begin{figure}
    \centering
    \includegraphics[width=.75\columnwidth,trim=470 0 470 0,clip]{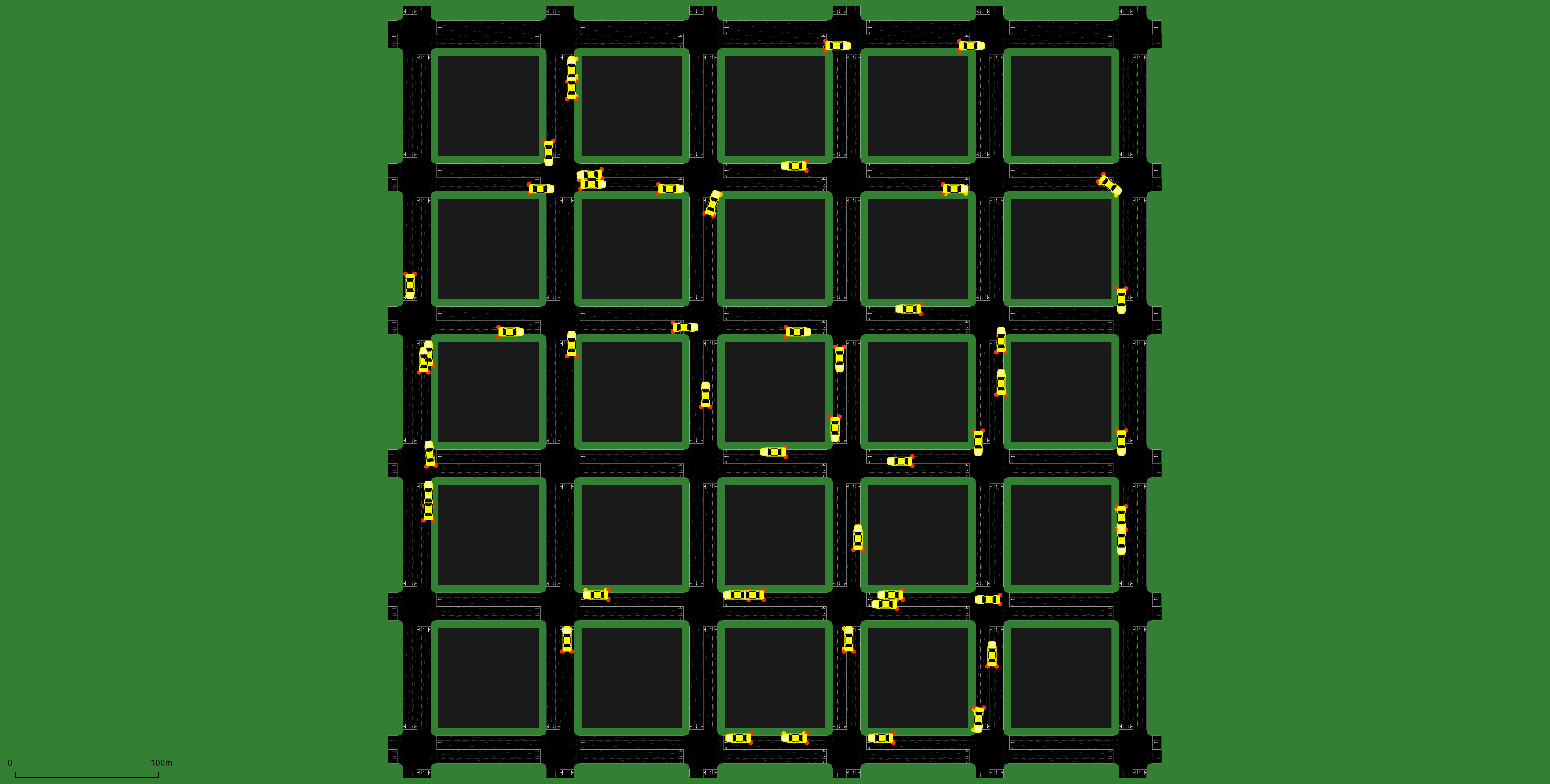}
    \caption{Manhattan grid with randomly distributed vehicles}%
    \label{fig:scenario}
\end{figure}

We focus on an urban simulation environment and, for simplicity, chose a \SI{550}{\m} $\times$ \SI{550}{\m} Manhattan grid scenario (see \cref{fig:scenario}).
The scenario contains \num{200} vehicles that depart at random positions and follow random trips.
Vehicles are transmitting beacons such as \acp{CAM} via \p{} at a static beacon rate.
In our simulation, we are able to switch off the attenuation effect of buildings by disabling the \emph{obstacle shadowing}~\cite{sommer2011computationally}.
Furthermore, we modified the \ac{MAC}-layer queue to replace the most recent packet if the maximum queue size is reached and a new beacon was generated from the application layer.
Together with a queue size of \num{1}, this results in always transmitting the most recent data in the beacon \cite{baiocchi2017model}.

The data from the received packets including its generation and reception times is stored in a simple 1-hop neighbor table on every vehicle.
We are thus able to calculate the standard \ac{AoI} for a given link by using this time stamp of the last successfully received update.
Whenever a new beacon from vehicle $j$ is received by vehicle $i$, the data as well as the time stamp is updated and we record the \ac{PAoI} as the current \ac{AoI} value of this link at $i$.
Similarly, we calculate the weighting coefficient according to \cref{eq:weighting_coefficient_omega} for a particular link upon successful packet reception by using the sender's position from within the packet.
Thus, these values are only calculated upon reception of at least the second beacon.
From our simulation, we obtain, on average, a total of \num{1500} samples for \ac{PAoI} and corresponding target \ac{AoI} per vehicle and simulated beacon rate, which we are going to use for the following results.

%

\subsection{Impact of Link Distance on Standard AoI}%
\label{sec:distance_peak_aoi}

In order to underline the issues with the standard \ac{AoI}, we first have a look at the impact of the link distance on the \ac{AoI}.
The \ac{PAoI}, as defined in \cref{eq:p_aoi}, is influenced by the beacon rate (cf.\ $\mathbb{E}\{Y\}$), system delay (cf.\ $\mathbb{E}\{T\}$), and probability of successful packet reception (cf.\ $\frac{1}{P_{sd}}$).
Thus, even if multiple vehicles use the same beacon rate for beacon transmission, the observed \ac{PAoI} metrics can be very different due to the effects of the wireless communication channel, especially over large distances.
Since the longest possible distance between two vehicles in our scenario is only about \SI{780}{\m}, we can neglect the system delay as an influencing factor.

The probability for successful reception of a packet, however, has to be considered.
It depends on the \ac{SNIR}, which is, among others, influenced by scenario-related effects such as attenuation of the signal as well as interference from other vehicles.
In our scenario, the signal is attenuated by free-space path loss, which weakens its strength proportional to the link distance.
Also, at large distances, hidden nodes can introduce additional interference and collisions, which further degrades the \ac{SNIR}.
At some point, a packet cannot be received successfully anymore and the \ac{AoI} of the corresponding link increases further until the next successful reception.
Therefore, the link distance can have a huge impact on the \ac{PAoI}, especially for far away vehicles.

\begin{figure}
    \centering
    \includegraphics[width=\columnwidth]{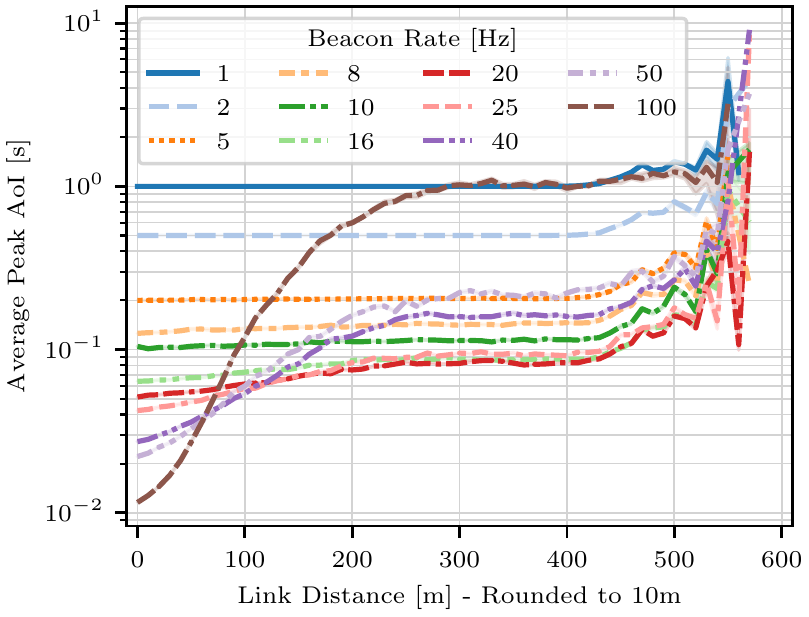}
    \vspace{-1.5em}
    \caption{Average \ac{PAoI} based on the link distance (rounded to \SI{10}{\m}) with different beacon rates}%
    \label{fig:distance_peak_aoi}
\end{figure}

\Cref{fig:distance_peak_aoi} shows the average \ac{PAoI} per link distance (rounded to \SI{10}{\m}) with different beacon rates that we obtained from the simulation.
Indeed, we see that the link distance has an impact on the \ac{PAoI} according to our hypothesis described previously.

For small link distances (less than \SI{50}{\metre}) and low beacon rates (e.g., \SI{1}{\hertz}), the observed \ac{PAoI} closely follows the beacon interval (i.e., the multiplicative inverse of the beacon rate) as the signal distortion due to the impact of the wireless communication channel is minimal.
However, at higher beacon rates (e.g., \SI{16}{\hertz}), the effect is increased and becomes visible more clearly.
Latest at roughly \SI{400}{\m}, we start to see a massive increase in \ac{PAoI}, which is way above the beacon interval.
Here, the probability for a successful reception of a packet is so low that many updates are lost and the \ac{PAoI} increases a lot.
For very high rates (e.g., \SI{40}{\hertz} and above), the \ac{PAoI} already steeply rises at even low distances of below \SI{100}{\m}.
As a result, we see that, even when using the same beacon rate, two links can have a very different \ac{PAoI} because of different node distances.
Hence, the freshness of the information from close vehicles is typically much better than from those far away, as expected.

%

\subsection{Combined Standard AoI}%
\label{sec:combined_aoi_standard}

Consider an arbitrary cooperative driving application that requires regular updates from surrounding vehicles (e.g., \ac{ICA}, cf.\ \cref{sec:spatial_model}).
This application will likely define a target \ac{AoI} (i.e., maximum allowed \ac{AoI}) that is required by the application to work successfully and reliably.
In order to determine whether certain information is fresh enough, the average \ac{PAoI} can be evaluated against this requirement.
A typical update interval requirement that is often found in literature is \SI{100}{\ms} \cite{ploeg2011design,heinovski2019modeling}.
Following the observation from the previous section, far away vehicles will always suffer from a weakened \ac{SNIR} and thus have a higher \ac{PAoI} compared to vehicles that are close.
Thus, when combining the \ac{PAoI} values from all surrounding vehicles using \cref{eq:p_aoi_neighbors}, the far away vehicles will increase the average and thus distort the view on the overall information freshness.

\begin{figure}
    \centering
    \includegraphics[width=\columnwidth]{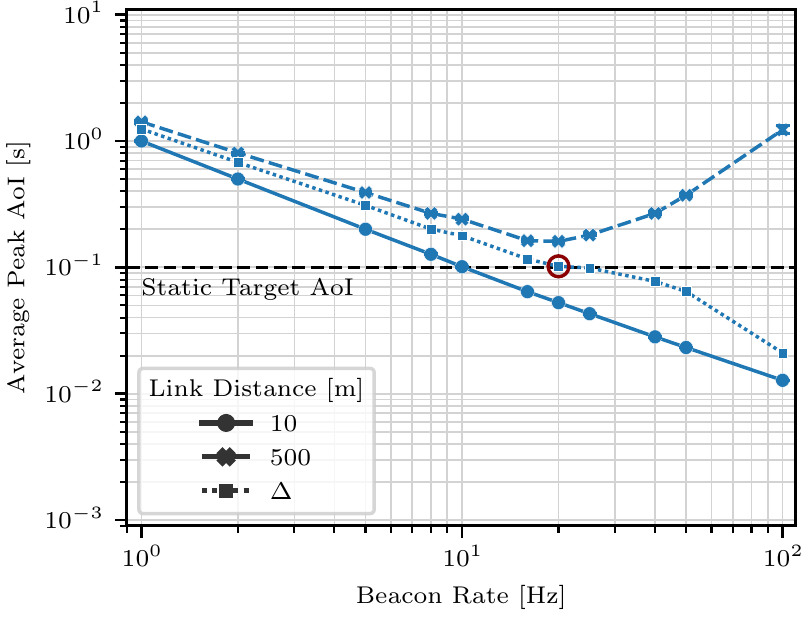}
    \vspace{-1.5em}
    \caption{%
        Average \ac{PAoI} plotted for all as well as separated for two specific node distances (i.e., short and long).
        The red circle indicates the intersection of the average \ac{PAoI} with the (static) target \ac{AoI} of \SI{100}{\ms}.
    }%
    \label{fig:paoi_target_2links}
\end{figure}

\Cref{fig:paoi_target_2links} shows the average standard \ac{PAoI} of short (i.e., \SI{10}{\m}) and long (i.e., \SI{500}{\m}) distance links for several beacon rates.
It also shows a static target \ac{AoI} of \SI{100}{\ms} as well as the average of the \ac{PAoI} values from the two links (cf.\ \cref{eq:p_aoi_neighbors}).
%
As expected, the average \ac{PAoI} of the short link distance (i.e., \SI{10}{\m}) continuously decreases proportional to the increasing beacon rate.
Here, the minimum value, which indicates the best information freshness, is reached at the highest simulated beacon rate (i.e., \SI{100}{\hertz}.
At this distance, the static target \ac{AoI} of \SI{100}{\ms} is already reached at a beacon rate close to \SI{10}{\hertz}.
This is expected, since \SI{10}{\hertz} is the multiplicative inverse of the target \ac{AoI} and the \ac{PAoI} is not distorted at these short distances (cf.\ \cref{sec:distance_peak_aoi}).

When looking at the long link distance (i.e., \SI{500}{\m}), the situation is different:
First, the average \ac{PAoI} is decreasing similarly to the short link distance, following the increase of the beacon rate.
But it never reaches the target \ac{AoI} of \SI{100}{\ms}.
Instead, after reaching its minimum at \SI{20}{\hertz}, the average \ac{PAoI} increases when beacons are transmitted at higher rates.
That is due to effects of the wireless communication channel described in \cref{sec:distance_peak_aoi}.

When looking at the combination (i.e., average) of the two link distances, we can observe some interesting effects as well.
First, the average \ac{PAoI} value decreases as expected, but when increasing the beacon rate further, also its value decreases further, thus reaching the static target \ac{AoI} of \SI{100}{\ms} at roughly \SI{20}{\hertz} (red circle).
The continuous decrease even at higher beacon rates (i.e., above \SI{20}{\hertz}) is due to less received packets for the long distance link.
Thus, the combined \ac{PAoI} contains a lot more low values which have been observed from the short link distance.

When we now compare the beacon rates at which the static target \ac{AoI} of \SI{100}{\ms} is met, we see that twice the beacon rate of the short distance links is required when combining the \ac{PAoI} metrics from both distances.
Hence, in order to keep the combined freshness of the information from all surrounding vehicles below the given \ac{AoI} requirement, beacons need to be transmitted at a higher rate than required for close vehicles only.
If vehicles now have different levels of relevance to the application, e.g., close vehicles are more important than far ones (cf.\ \cref{sec:spatial_model}), the non-relevant vehicles (i.e., the far ones) will weaken the perceived combined information freshness.
In order to meet the \ac{AoI} requirement, all vehicles need to transmit their beacons at a higher rate, which leads to unnecessary transmissions and channel load.

%

\subsection{Combined Weighted AoI}%
\label{sec:combined_aoi_spatial}

In order to cope with the issues of the standard \ac{AoI} (i.e., effects of the wireless communication channel and equal importance of all nodes), we now apply the proposed spatial model from \cref{sec:spatial_model} to the \ac{AoI}.
Using \cref{eq:weighting_coefficient_omega}, we calculate the weighting coefficient $\omega$ for every \ac{PAoI} value that is observed for an arbitrary link between two vehicles $i,j$, producing a \emph{weighted \ac{PAoI}}.
Additionally, we also apply the weighting coefficients to the static target \ac{AoI} of \SI{100}{\ms} on a per link bases, producing a \emph{weighted target \ac{AoI}}.
In this section, we use one exemplary parameterization (i.e., $\alpha = 0, \beta = 0.01$) of our spatial model that uses only the distance between vehicles for calculating the weighting coefficient.
This focuses on the issue described in \cref{sec:distance_peak_aoi}.

\begin{figure}
    \centering
    \includegraphics[width=\columnwidth]{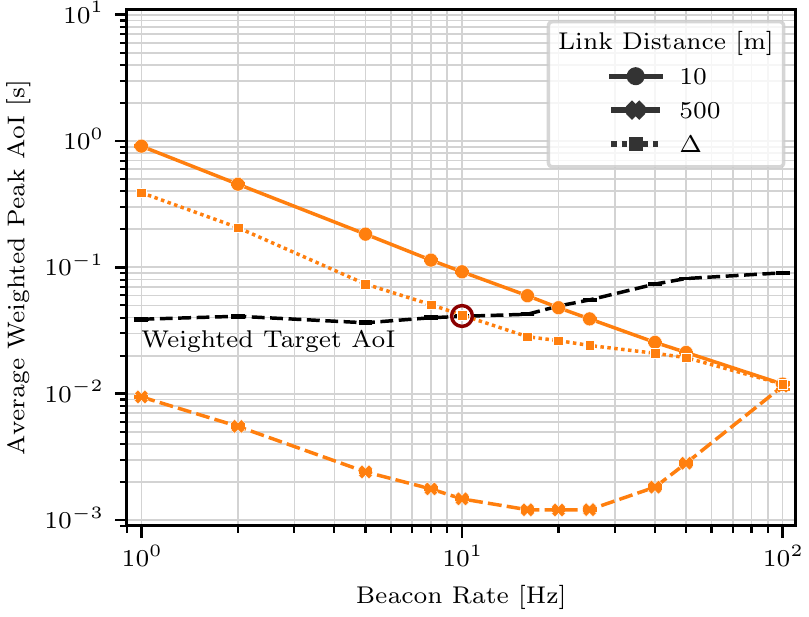}
    \vspace{-1.5em}
    \caption{%
        Average weighted \ac{PAoI} for two specific link distances (i.e., short and long) and their combination (i.e., average).
        The red circle indicates the intersection of the combined \ac{PAoI} with the weighted target \ac{AoI}.
    }%
    \label{fig:weighted_aoi_2links}
\end{figure}

\Cref{fig:weighted_aoi_2links} shows the average weighted \ac{PAoI} of short (i.e., \SI{10}{\m}) and long (i.e., \SI{500}{\m}) distance links over several beacon rates.
It also shows an average weighted target \ac{AoI} as well as the average \ac{PAoI} values from the two links, which can be used by application as a view on the overall information freshness.
Since we selected fixed distances, the calculated weighting coefficient will be the same for all links of the same distance.
The resulting average weighted \ac{PAoI} is just a multiplication of the average standard \ac{PAoI} from \cref{fig:paoi_target_2links} with a constant factor and thus follows a similar trend.

Due to the selected parameterization of our spatial model, a high (i.e., close to \num{1}) and a low (i.e., close to \num{0}) weighting coefficient is calculated for values of the short and long distance links, respectively.
As a result, the average \ac{PAoI} for the short distance links is very close to the one of the standard \ac{AoI} from \cref{fig:paoi_target_2links}, whereas it is reduced a lot for the long distance links.
Therefore, and due to less observations for the long distance links in general, the combination (i.e., average) of all \ac{PAoI} values from both distances is, in comparison to \cref{fig:paoi_target_2links}, much closer to the average \ac{PAoI} of the short distance links.
Thus, the overall view on the information freshness is not distorted much by the long distance link, which is (in our parameterization) less relevant.

The average weighted target \ac{AoI} is constructed by combining all weighted target \ac{AoI} values from the two link distances, similarly to the average weighted \ac{PAoI}.
Since the same weighting coefficients are applied to the \ac{PAoI} and the target \ac{AoI}, the target faces similar effects:
For the short distance, the target is close to the static target \ac{AoI} of \SI{100}{\ms}, whereas for the long distance, it is close to \num{0} due to the weighting coefficient being close to \num{0}.
The average weighted target \ac{AoI} thus is close to the static target \ac{AoI} of \SI{100}{\ms} as it is mostly influenced by short distance links.
Note that the individual weighted target \ac{AoI} for both distances is constant for all beacon rates as the link distance does not change and the beacon rate is not considered when calculating the weighting coefficient.
The average weighted target \ac{AoI}, in contrast, is not constant due to the increasing number of lost packets and thus less values for the long link distance with high beacon rates.
The average therefore tends towards the value of the short link distance, when using a beacon rate $\geq\SI{20}{\hertz}$.

When comparing the average \ac{PAoI} with the target \ac{AoI}, we see that now both link distances as well as their combination intersect with the target \ac{AoI} at some point.
The short distance links meet the target at a beacon rate close to \SI{20}{\hertz}.
The average \ac{PAoI} of the long distance links is below the weighted target \ac{AoI} even for all beacon rates.
This is due to the average weighted target \ac{AoI} mainly begin influenced by the short distance links, thus, tending towards the static target \ac{AoI} of \SI{100}{\ms}.
Additionally, the weighting coefficient for the long distance links are close to \num{0}.
The combined weighted \ac{PAoI} reaches the weighted target \ac{AoI} at a beacon rate close to \SI{10}{\hertz}, as indicated by the red circle.
This is a smaller beacon rate than required for only the short distance links due to the impact of the long distance links on the average.
However, we actually need to compare this situation with the values from using the standard \ac{PAoI} and the static target \ac{AoI} in \cref{sec:distance_peak_aoi}:
With the spatial model, we only need half of the previous beacon rate to reach the target \ac{AoI} when combining all link distances.
Using our model allows to focus on timely updates of relevant vehicles for meeting a given \ac{AoI} requirement instead, which saves channel resources.

\subsection{Ratio of Links Reaching the Target AoI}%
\label{sec:ratio_comparison}

\begin{figure}
    \centering
    \includegraphics[width=\columnwidth]{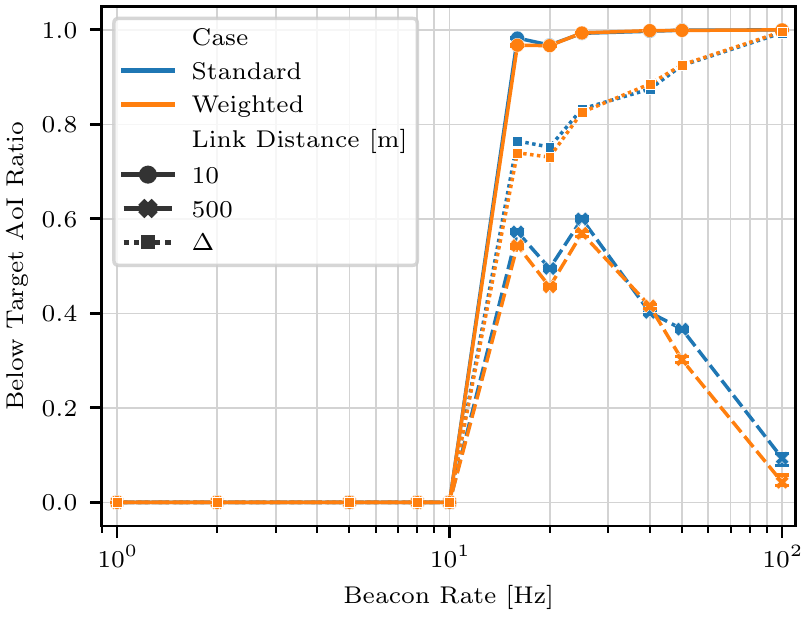}
    \vspace{-1.5em}
    \caption{%
        Ratio of vehicle updates that are below the target \ac{AoI} for the standard case (blue) and when applying our spatial model (orange)
        -- two specific link distances (i.e., short and long) and their combination (i.e., average).
    }%
    \label{fig:below_aoi_target_2links}
\end{figure}

The weighted target which we introduced in \cref{eq:target_link} allows interpreting the weighted \ac{PAoI} in the way we interpret the standard one.
To illustrate the validity of our approach,
\cref{fig:below_aoi_target_2links} comparatively depicts the ratio of \ac{PAoI} values fulfilling the condition in \cref{eq:target_link_leq} and the case using the standard approach, i.e., $\Delta_{j,i} \leq T_{j,i}$ without using the weighting coefficient.
Without the spatial model, the target \ac{AoI} is static at \SI{100}{\ms}, whereas when applying the spatial model, it is calculated by using the weighting coefficient (cf.\ \cref{eq:target_link}).


The ratio for both cases is at \num{0} for all beacon rates $\leq \SI{10}{\hertz}$.
This is expected since the target \ac{AoI} of \SI{100}{\ms} cannot be reached when the inter-arrival time of the beacons is larger than this value.
When increasing the beacon rate further (above \SI{10}{\hertz}), all ratios are increasing as well.
The ratios for the short link distance almost immediate reach \num{1} and stay there since these links have a very good \ac{SNIR} and thus almost all transmitted beacons are received successfully, leading to a small \ac{PAoI}.
For the long link distance, the ratio grows as well but not as strongly as for the close links.
Again, this is due to the weighted \ac{PAoI} being impacted by the large distance of the link (cf.\ \cref{fig:distance_peak_aoi}), thus leading to many \ac{PAoI} values being above the target \ac{AoI}.
Beyond \SI{25}{\hertz} beacon rate, the ratio for the long link distance decreases again due to a high \ac{PAoI} (cf.\ \cref{fig:weighted_aoi_2links}).

As expected, the combination of both link distances lies in between the short and the long distance links.
After reaching roughly \num{0.75} at \SI{16}{\hertz}, its value increases until it reaches almost \num{1} at \SI{100}{\hertz}.
This is due to the high number of lost packets for the long distance links, which leads to a value similar to the value of the short link distance.

It is visible that the ratio is very similar for both cases (i.e., standard and weighted).
This reflects that the perceived status update on the network is the same irrespective of the weighting coefficients.
Thus, using the formulation in \Cref{eq:target_network_leq} will not introduce any artifact on the perceived network status; instead, it will just emphasize the relevant links.

%

\subsection{Weighted Network AoI}%
\label{sec:network_aoi}

So far, in order to show how our spatial model can influence the perceived average \ac{PAoI} when combining different links, we have been looking at two specific link distances only.
Now, we combine all available links from within the simulation scenario to analyze the average \ac{PAoI} of the entire network.
Again, we are evaluating the freshness of the information by comparing the average \ac{PAoI} against the \ac{AoI} requirement.
This time, however, the goal is to determine the overall information quality of the entire network.

\begin{figure}
    \centering
    \includegraphics[width=\columnwidth]{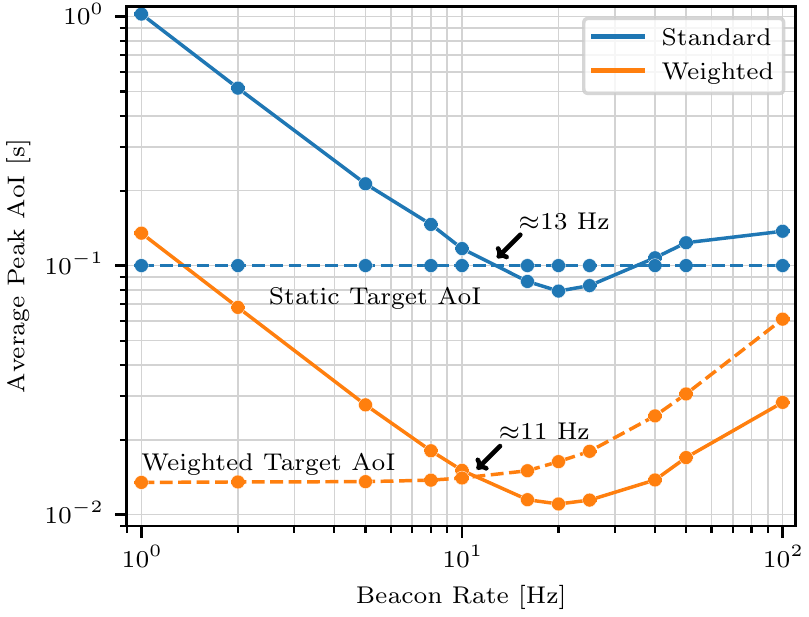}
    \vspace{-1.5em}
    \caption{%
        Average \ac{PAoI} (solid) and corresponding target \ac{AoI} (dashed) of the entire network for the standard case (blue) and when applying our spatial model (orange).
        The arrows indicate the intersection with the corresponding target \ac{AoI}.
    }%
    \label{fig:network_peak_aoi_target}
\end{figure}

\Cref{fig:network_peak_aoi_target} shows the average network \ac{PAoI} (solid, cf.\ \cref{eq:p_aoi_network}) as well as the corresponding network target \ac{AoI} (dashed) for the standard case (blue) and when applying our spatial model (orange).
We use the same parameterization of the spatial model as we did already in \cref{sec:combined_aoi_spatial} (i.e., $\alpha = 0, \beta = 0.01$).
The standard \ac{PAoI} (blue) behaves as expected.
It follows the beacon rate inversely proportional as the interval between beacons decreases with higher beacon rates.
It intersects with the static target \ac{AoI} of \SI{100}{\ms} at approximately \SI{13}{\hertz} (not simulated) and reaches its minimum value at \SI{20}{\hertz}.
Since the average \ac{PAoI} here contains the values from all available link distances (i.e., \SIrange{0}{600}{\m}), there is indeed a minimum, which we can not observe in \cref{fig:paoi_target_2links}.
This is due enough successfully received packets with high \ac{PAoI} values (mostly from long distance links) such that they can weaken the perceived average \ac{PAoI}.

When looking at the weighted case (orange), the situation is similar but the absolute values of the average \ac{PAoI} and target \ac{AoI} are smaller due to the applied weighting coefficient.
Note, that the minimum \ac{PAoI} value is achieved at the same beacon rate in both cases, which underlines the applicability of our model without distorting the standard interpretation of the \ac{AoI}.
Similar to \cref{fig:weighted_aoi_2links}, the target is calculated by combining all available individual target \ac{AoI} values using the average function (see \cref{eq:target_network}).
It is almost constant and lower in comparison to the static target \ac{AoI} at beacon rates $\leq \SI{10}{\hertz}$ due to many medium and long distance links that have a small weighting coefficient.
At these low beacon rates, the packets from large distances can still be received successfully.
When increasing the beacon rate beyond \SI{10}{\hertz}, analog to the average \ac{PAoI}, the number of lost packets for medium and long distance links increases and the average weighted target \ac{AoI} therefore tends towards the value of the short link distances.
In the weighted case, the intersection of the average \ac{PAoI} with the target happens already at approximately \SI{11}{\hertz} (not simulated), which indicates that this beacon rate is high enough to achieve the required \ac{AoI} of the entire network on average.
This approximated beacon rate is roughly \SI{2}{\hertz} lower compared to the standard case.
Using our model thus allows to save channel resources by focusing on timely updates of relevant vehicles for meeting a given \ac{AoI} requirement.

%

\subsection{Impact of Model Parameters}%
\label{sec:eval_model_parameters}

In our simulative results, so far we have only looked at one exemplary parameterization (i.e., $\alpha = 0, \beta = 0.01$) of our spatial model that uses only the distance between vehicles for calculating the weighting coefficient.
In fact, we used a very strict value for the distance parameter $\beta$, which was favouring very short link distances.
In general, more relaxed configurations will lead to a higher weighting coefficient, thus, producing higher \ac{PAoI} (cf.\ \cref{sec:analytical_results}), especially for vehicles far away from the front of the receiver (i.e., in distance and orientation).
Thus, we now compare the resulting average \ac{PAoI} as well as the target \ac{AoI} for different parameterizations of our spatial model.

\begin{figure}
    \centering
    \includegraphics[width=\columnwidth]{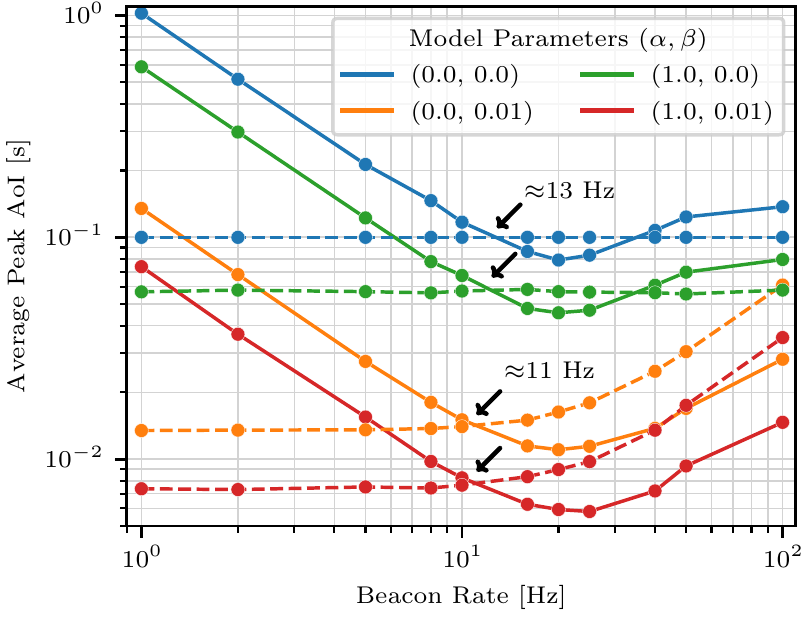}
    \vspace{-1.5em}
    \caption{%
        Average \ac{PAoI} (solid) and corresponding target \ac{AoI} (dashed) of the entire network for different configurations of our spatial model.
    }%
    \label{fig:paoi_spatial_parameters}
\end{figure}

\Cref{fig:paoi_spatial_parameters} shows the average \ac{PAoI} (solid) and corresponding target \ac{AoI} (dashed) of the entire network for different configurations of our spatial model.
Here, we focus only on \num{4} different configurations:
\begin{enumerate}
\item $\alpha = 0, \beta = 0$, which reflects the standard case by always using a weighting coefficient of \num{1} (blue),
\item $\alpha = 0, \beta = 0.01$, which only focuses on the distance between vehicles for determining their relevance (orange, see previous sections),
\item $\alpha = 1, \beta = 0$, which only focuses on the orientation (angle) between vehicles for determining their relevance (green),
\item $\alpha = 1, \beta = 0.01$, which uses orientation (angle) and distance between vehicles for determining their relevance (red).
\end{enumerate}

As expected, all configurations that apply our spatial model (i.e., \numrange{2}{4}) result in a decrease of the average \ac{PAoI} and target \ac{AoI} by filtering less relevant vehicles.
We can observe, however, that using only the angle (2) results in a situation that is close to using the standard \ac{PAoI} and the static target \ac{AoI} (1).
In contrast, using only the distance (3) results in a situation that is close to using angle and distance together (4).
Configuration which behave similarly also have a similar beacon rate at which the average \ac{PAoI} is intersecting with the target \ac{AoI}, i.e., approximately \SI{13}{\hertz} vs. \SI{11}{\hertz} (not simulated).
This shows that the distance has more impact than the angle when calculating the weighting coefficient, which is in line with the theoretical results from \cref{sec:analytical_results}.
This is due to the effects of the wireless communication channel (cf.\ \cref{sec:distance_peak_aoi}), which impact the \ac{PAoI} quite heavily.
Also, vehicles within the scenario are distributed in space rather than in the close surroundings of single vehicles.


%

\subsection{Impact of the Scenario}%
\label{sec:eval_scenario}


Until now, we had disabled all movement of the vehicles as well as the attenuation of the wireless signal by buildings in our Manhattan grid simulation scenario.
We did this such that we can clearly observe the effects of our spatial model.
It is obvious that this is not realistic and that enabling both of these will influence the wireless channel available to the vehicles.

When considering buildings, the wireless signal will be attenuated, leading to unsuccessful transmissions for links with medium and long distances, thus increasing their \ac{AoI} (if a message is received at all).
In contrast, the \ac{AoI} for short link distances will likely be improved due to less interference from far-away vehicles.
We expect this effect in particular within our Manhattan grid scenario as \ac{LOS} between vehicles exists rarely, mainly only for vehicles on the same road or within the same road corridor.
Additionally, every building's wall attenuates the signal of \ac{NLOS} links even more.
Thus, the interference domain for every receiving vehicles is reduced a lot in comparison to the same scenario without buildings.

When considering mobility, the distance as well as the signal quality of the links will change over time as vehicles drive around.
A former short distance link between two vehicles with a good signal quality can become a medium or even long distance one with worse quality (and \ac{AoI}), and the other way around.
However, in a free-space scenario, we expect these changes and the differences between vehicles to be limited and the link quality to stay rather equal as vehicles can only drive a certain distance in a given time and continuously have a \ac{LOS} connection.

When considering the real-world scenario, i.e., combining attenuation by buildings and vehicle mobility, both of the aforementioned effects are mixed:
At one point in time, an arbitrary link can have a good quality with a low \ac{AoI}, but at a different point in time, the same link can have a very high \ac{AoI} or be blocked completely because vehicles changed their position.
The opposite can happen as well: vehicles that did not have a \ac{LOS} connection before, leading to bad link quality and \ac{AoI} values, can suddenly be on the same road corridor leading to a good \ac{LOS} connection.

In order to show the impact of the three aforementioned cases in comparison to the previously used simulation scenario (i.e., free-space, no mobility), we repeated the previous simulation study with the following four scenarios:
\begin{enumerate}
\item \emph{Freespace}, a scenario without buildings but with vehicle mobility
\item \emph{Freespace static}, a scenario without buildings and without vehicle mobility (scenario from above)
\item \emph{Manhattan}, a scenario with buildings and with vehicle mobility (real-word scenario)
\item \emph{Manhattan static}, a scenario with buildings but without vehicle mobility
\end{enumerate}
For these simulations, we set the simulation time to \SI{30}{\s}, such that the vehicle mobility can have an effect.
In the remainder of this section, we only show results for the configurations $(0.0, 0.0)$ (standard case) and $(1.0, 0.01)$ (strictest weighted case, see \cref{fig:paoi_spatial_parameters}).

\begin{figure}
    \centering
    \subfloat[\Acf{CBR}. The dashed line indicates the practical maximum at \num{80}\% (cf.\ \ac{CSMA/CA} in \p{}).]{\includegraphics[width=\columnwidth]{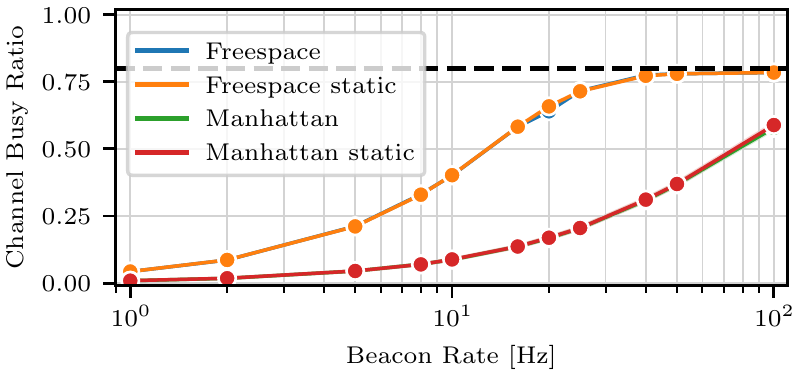}\label{fig:all-cbr}}\\
    \subfloat[\Acf{BRR}]{\includegraphics[width=\columnwidth]{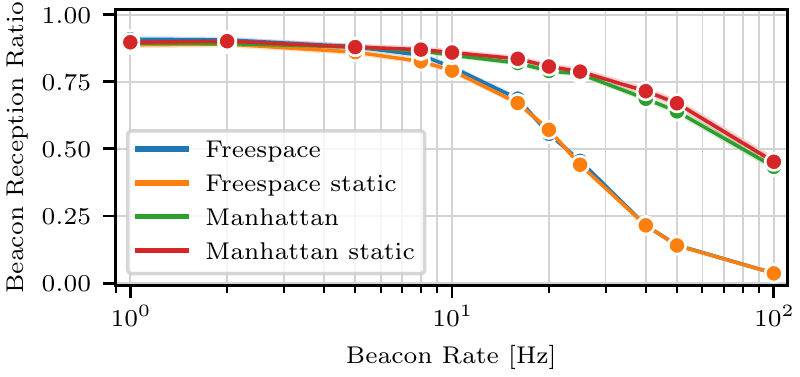}\label{fig:all-brr}}
    \vspace{-.5em}
    \caption{%
        Key networking metrics for different scenarios.
    }%
    \label{fig:all_networking_metrics}
\end{figure}

In order to visualize the impact of the scenarios on networking metrics, \cref{fig:all-cbr,fig:all-brr} shows simulation results for \ac{CBR} and \ac{BRR}, respectively.
Values are recorded for all \num{200} vehicles during the entire simulation time and are time-weighted averages, which we aggregate to overall averages.
Since the spatial model does not impact these metrics, we do not visualize the different cases (standard vs. weighted).
Following our expectations from above, buildings in the scenario have a huge impact on the signal quality and interference domain of vehicles and thus on the channel load.
With buildings (\emph{Manhattan} and \emph{Manhattan static}), the wireless signals are heavily attenuated, leading to reduced absolute values and a less steep increase of the \ac{CBR} in comparison to the scenarios without buildings.
Even at high beacon rates (above \SI{40}{\hertz}), these scenarios still have a relatively low \ac{CBR} whereas both scenarios without buildings reach the practical maximum value of \num{80}\% (cf.\ \ac{CSMA/CA} in \p{}).
The results for the \ac{BRR} are in line with the observed \ac{CBR}:
We observe a lot of collisions and thus lost beacons at high beacon rates, especially in the scenarios without buildings where the interference domain is large.
In contrast to buildings, the vehicle mobility has only little impact on the channel load (see \emph{static} vs. non-\emph{static} scenarios) as the signal quality and interference domain does not change much by movement.
This is inline with our expectations from above and can be observed in the \ac{BRR} as well.

\begin{figure}
    \centering
    \includegraphics[width=\columnwidth]{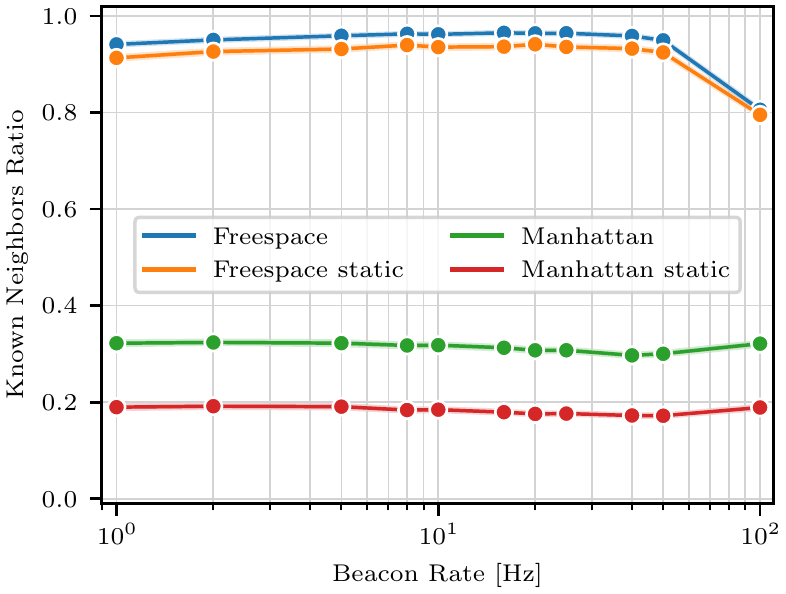}
    \vspace{-1.5em}
    \caption{%
        Average ratio of known neighbors (i.e., other vehicles from which at least \num{1} beacon was received) for different scenarios.
    }%
    \label{fig:all_knowledge}
\end{figure}

In order to analyze vehicles' knowledge of the scenario, we report the ratio of known neighbors (i.e., other vehicles from which at least \num{1} beacon was received) in \cref{fig:all_knowledge}.
Again, we can observe the impact of the buildings, which block the signal of a lot of transmissions:
While almost all other vehicles (neighbors) in the scenario are known (i.e., at least \num{1} beacon was received) within \emph{Freespace} and \emph{Freespace static}, only some are known within \emph{Manhattan} and \emph{Manhattan static}.
Here, the transmissions are blocked by the buildings, leading to no received beacons and thus no knowledge about a lot of other vehicles in the scenario.
The values are (almost) constant over all (but the highest) beacon rates for all scenarios as the signal quality and interference domain does not change much due to vehicle movement and receiving at least \num{1} beacon during the entire simulation is possible even with high channel load (see \cref{fig:all-cbr}).
This, of course, changes when requiring regular updates as packets are lost and updates are missed due to collisions.
Following our expectations from above, mobility has only little impact for the scenarios without buildings.
However, we can observe that it has indeed an impact and can increase the knowledge if the scenario contains buildings (\emph{Manhattan} vs. \emph{Manhattan static}).
Here, mobility helps achieving links and thus knowledge about other vehicles by moving either into \ac{LOS} or at least increasing signal quality for \ac{NLOS} paths.

\begin{figure}
    \centering
    \includegraphics[width=\columnwidth]{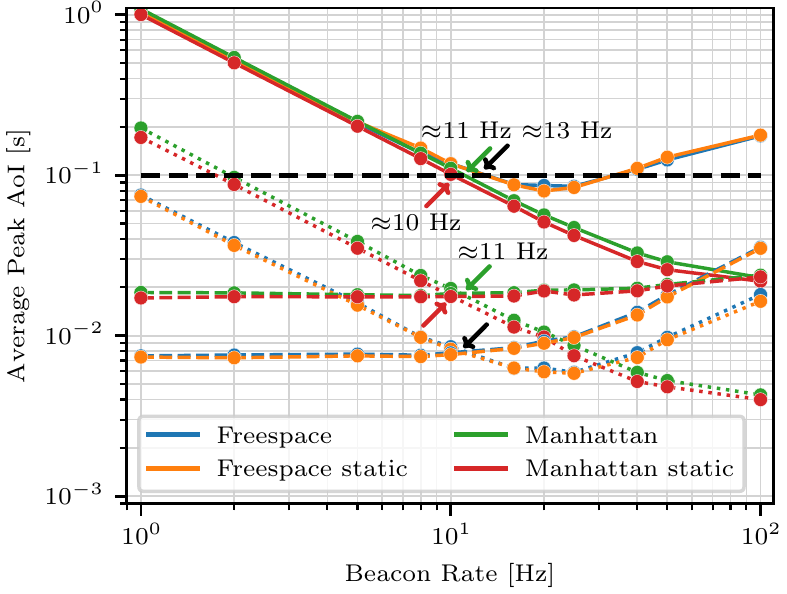}
    \vspace{-1.5em}
    \caption{%
        Average \ac{PAoI} (solid and dotted lines) and corresponding target \ac{AoI} (dashed lines) of the entire network for all scenarios).
        The solid lines show the standard \ac{PAoI} wheres the dotted lines show the weighted \ac{PAoI} with an exemplary configuration of $\alpha = 1.0, \beta = 0.01$.
        The corresponding target \ac{AoI} is shown by the dashed lines (black $=$ standard, color $=$ weighted).
    }%
    \label{fig:all_network_peak_aoi_target}
\end{figure}

\Cref{fig:all_network_peak_aoi_target} shows the average network \ac{PAoI} (solid and dotted lines) as well as the corresponding network target \ac{AoI} (dashed lines) for all scenarios.
Here, the solid lines correspond to the standard \ac{PAoI} (no spatial model) whereas the dotted lines correspond to the weighted \ac{PAoI} ($\alpha = 1.0, \beta = 0.01$).
The latter one needs to be evaluated against the weighted the target \ac{AoI} (colored dashed lines) whereas the standard case is evaluated against the target \ac{AoI} of \SI{100}{\ms} (black dashed line).

Following the previously described effects, the spatial model reduces the values for \ac{PAoI} and target \ac{AoI} in all scenarios.
The results for the scenarios without buildings (\emph{Freespace static} and \emph{Freespace}) are in line with \cref{fig:network_peak_aoi_target} for both case:
the \ac{PAoI} follows the beacon rate, meets the target at \SI{13}{\hertz} and \SI{11}{\hertz} for the standard and weighted case, respectively, and increases again at higher beacon rates, which leads to a minimum at \SI{20}{\hertz} (for the standard case).
Vehicle mobility here has almost no impact since all vehicles already receive all signals and movement does not change the interference domain.

For scenarios with buildings (\emph{Manhattan} and \emph{Manhattan static}), we can observe some interesting effects.
Both scenarios have smaller standard \ac{PAoI} values because of the smaller interference domain and higher beacon reception ratio (see \cref{fig:all-brr}) due to buildings.
Accordingly, the channel will be saturated only at very high beacon rates, which leads to a minimal \ac{PAoI} value at the highest simulated beacon rate of \SI{100}{\hertz}.
Here, the \ac{PAoI} behaves similarly to the short distance links from \cref{sec:combined_aoi_standard}.
Following the smaller beacon reception ratio from \cref{fig:all-brr}, the vehicle mobility in the \emph{Manhattan} scenario leads to slightly higher \ac{PAoI} values.
This is due to the fact that mobility achieves knowledge about \emph{new} neighbors but increases the \ac{AoI} for \emph{old} neighbors at the same time.

When applying the spatial model, the \ac{PAoI} as well as the target \ac{AoI} is decreased as expected.
However, they are still higher than the values of scenarios without buildings because the weighting coefficient $\omega_{i,j}$ overall has larger values.
This is due to the current configuration of the spatial model ($\alpha = 1.0, \beta = 0.01$) and the overall smaller distances and limited set of angles for the links between vehicles due to the buildings (cf.\ \cref{fig:spatial_model}).
At the same time, the weighted target \ac{AoI} is almost constant for both scenarios with buildings, being in-line with the metric of known neighbors (\cref{fig:all_knowledge}).
The result of both of these effects is that the weighted \ac{PAoI} reaches its corresponding target \ac{AoI} at the same beacon rate ($\approx\SI{10}{\hertz}$) as in the standard case.
Thus, with this configuration of the spatial model, there is at least no visible benefit.

%

\subsection{Adaptive Beaconing}%
\label{sec:adaptive_results}

In this section, we analyze the behavior of our \ac{AoI}-based algorithm for adapting the beacon rate as well as the impact by the spatial model.
We use the \emph{Manhattan} scenario from \cref{sec:eval_scenario} with buildings and vehicle mobility and a target \ac{AoI} of \SI{100}{\ms}.
Vehicles start with an initial beacon rate of \SI{10}{\hertz}, which is the theoretical minimal beacon rate required to reach the target \ac{AoI}, assuming no propagation delay and packet collisions.
We simulate for \SI{30}{\s} but treat the first \SI{10}{\s} as warm-up period, during which the beacon rate achieves a steady-state.

\begin{figure}
    \centering
    \includegraphics[width=\columnwidth]{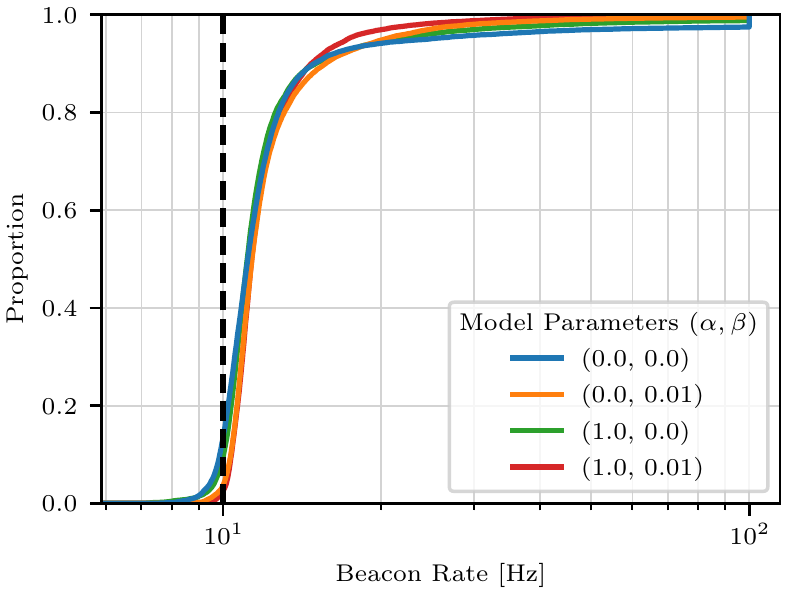}
    \vspace{-1.5em}
    \caption{%
        eCDF of observed beacon rate for the static target \ac{AoI} of \SI{100}{\ms} for different configurations of our spatial model.
        The dashed line indicates the theoretical minimal required beacon rate in order to reach the target \ac{AoI} (assuming no propagation delay and collisions).
    }%
    \label{fig:adaptive_beaconrate_ecdf}
\end{figure}
%

\Cref{fig:adaptive_beaconrate_ecdf} shows the beacon rates observed from all vehicles in form of an \ac{eCDF} for different configurations of our spatial model.
The adaptive algorithm leads to a beacon rate above the theoretical minimal beacon rate (\SI{10}{\hertz}) required to reach the target \ac{AoI} of \SI{100}{\ms} in most cases.
In contrast, only a few values are below the minimum rate, leading to too infrequent updates in all configurations.
However, the values are influenced by the configuration of our spatial model:
In general, we can observe that applying our spatial model leads to a better beacon rate selection (a better fit to the required beacon rate of \SI{10}{\hertz}).
This becomes more visible with stricter configurations (larger values of $\alpha$ and $\beta$, c.f\ \cref{sec:eval_model_parameters}).
The beacon rates are less distributed (steeper \ac{eCDF} curve) and closer to the theoretical minimal beacon rate.
E.g., the mean value is at \SI{12}{\hertz} and \SI{14}{\hertz} for the strictest configuration (red) and the standard case (blue).
In fact, the standard case has the most values below the minimal beacon rate of \SI{10}{\hertz} (\num{13}\%) and above \SI{20}{\hertz}, which is double the minimal beacon rate, (\num{6}\%).
In comparison, the strictest weighted case has only \num{2}\% and \num{3}\% for low and high values, respectively.
Similarly, the spatial model is able to reduce the amount of very large beacon rates:
$\approx$\SI{24}{\hertz} vs.\ $\approx$\SI{17}{\hertz} for the \num{95}th percentile, \SI{100}{\hertz} (the maximum possible beacon rate) vs.\ $\approx$\SI{23}{\hertz} for the \num{98}th percentile, and \SI{100}{\hertz} vs.\ $\approx$\SI{33}{\hertz} for the \num{99}th percentile.
The observed \ac{CBR} follows accordingly with a mean of \num{12}\% vs.\ \num{10}\% and a \num{99}th percentile of \num{18}\% vs.\ \num{16}\% for the standard case (blue) and the strictest model configuration (red).
Overall, using the spatial model avoids outliers in both directions (too little and too high values) during beacon rate selection, which prevents too infrequent updates and high channel usage.


\begin{figure}
    \centering
    \includegraphics[width=\columnwidth]{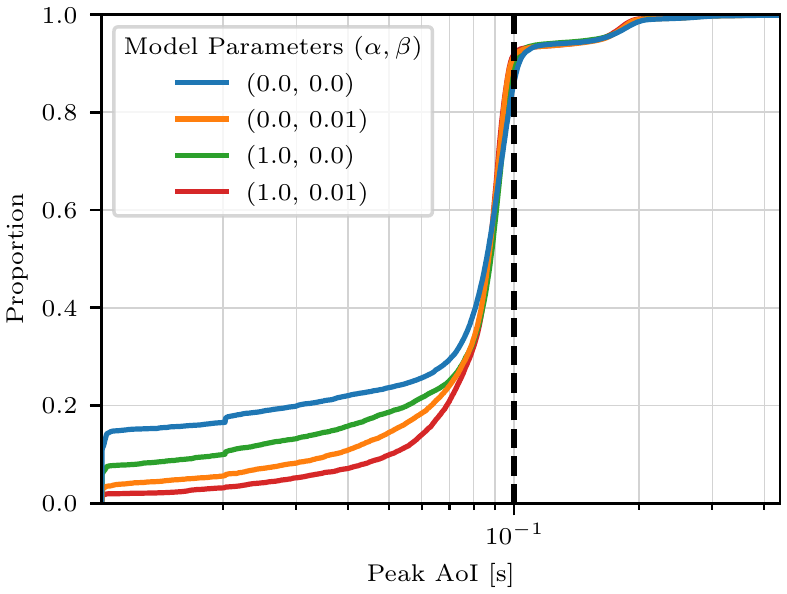}
    \vspace{-1.5em}
    \caption{%
        eCDF of observed \ac{PAoI} for the target \ac{AoI} of \SI{100}{\ms} (dashed line) for different configurations of our spatial model.
    }%
    \label{fig:adaptive_peak_aoi_ecdf}
\end{figure}
%
%

\Cref{fig:adaptive_peak_aoi_ecdf} shows the \ac{PAoI} measurements observed from all vehicles in form of an \ac{eCDF} for different configurations of our spatial model.
Since the \ac{AoI} is directly related to the beacon rate, the values follow the previously described effects (see \cref{fig:adaptive_beaconrate_ecdf}):
The stricter the spatial model configuration, the closer are the values to the target \ac{AoI} of \SI{100}{\ms}, which serves as the set-point for the \ac{PID} controller (see \cref{sec:pid_controller}).
Thus, the strictest configuration (red) has the largest but also closest \ac{PAoI} values with the least spread.
It has slightly higher mean \ac{PAoI} values (\SI{0.08}{\s} vs.\ \SI{0.09}{\s}) but leads to less deviation, e.g., \SI{0.057}{\s} vs.\ \SI{0.074}{\s} at the \num{25}th percentile and \SI{0.095}{\s} vs.\ \SI{0.092}{\s} at the \num{75}th percentile for the standard case (blue) and the strictest model configuration (red), respectively.

\begin{figure}
    \centering
    \includegraphics[width=\columnwidth]{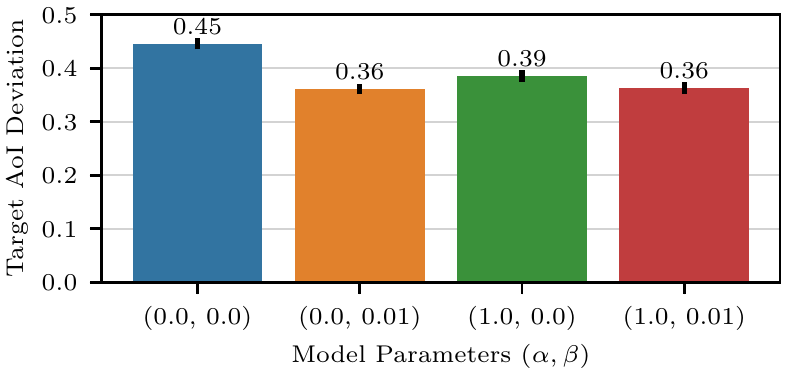}
    \vspace{-1.5em}
    \caption{%
        Average deviation (ratio) from target \ac{AoI} of \SI{100}{\ms} for different configurations of our spatial model.
    }%
    \label{fig:adaptive_diff_target}
\end{figure}

To visualize this effect more strongly, \cref{fig:adaptive_diff_target} shows the average deviation from target \ac{AoI} of \SI{100}{\ms} for different configurations of our spatial model.
Again, stricter model configurations lead to lower deviation, e.g., \num{67.5}\%, \num{45.7}\%, \num{38.4}\%, and \num{32.3}\% at the \num{75}th percentile.
In general, the overall performance of the system also increases when applying our spatial model:
The strictest configuration reduces the outliers with large \ac{PAoI} values from \SI{0.21}{\s} to \SI{0.19}{\s} at the \num{99}th percentile.
Thus, the ratio of observed \ac{PAoI} measurements that are below their respective target \ac{AoI} increases accordingly with stricter configurations (\num{86}\%, \num{89.9}\%, \num{91.5}\%, and \num{92.3}\%).
The remaining values above the target \ac{AoI} correspond mostly to beacon rates smaller than \SI{10}{\hertz} (see \cref{fig:adaptive_beaconrate_ecdf}).

\section{Discussion \& Remarks}%
\label{sec:discussion}

Throughout this work, we used a static target \ac{AoI} of \SI{100}{\ms}, since this value is often used in literature as a desired update interval for information used by cooperative driving applications.
This value, however, is arbitrary and can freely be configured dependent on the specific needs of the chosen application.
Our proposed spatial model is independent of the actual value for this target \ac{AoI}, since the same weighting coefficient is applied to both, the observed \ac{PAoI} values as well as the static target \ac{AoI}.
When modifying the value of the target \ac{AoI}, the intersection point with the average \ac{PAoI} will be shifted along the x-axes, leading to a different beacon rate that is sufficient for meeting the target.
When the target value is increased, this beacon rate decreases and vice versa.
In case both, a small target \ac{AoI} and a low beacon rate, is desired, it can be beneficial to use a stricter configuration of our spatial model (i.e., lager values for $\alpha$ and $\beta$).
This will impose a greater selectivity in the importance of vehicles and cope with effects of the wireless communication channel as described in \cref{sec:distance_peak_aoi}.

We chose to aggregate \ac{PAoI} values from multiple neighboring vehicles using the average function instead of some high percentile (e.g., \num{90}th or \num{99}th) due to multiple reasons.
The average \ac{PAoI} is a typical metric used in the field of \acl{AoI} \cite{yates2021age}.
When using high percentiles instead of an average in order to focus on neighbors with bad freshness (i.e., high \ac{PAoI}), the aggregated \ac{PAoI} value will be worse (i.e., higher) and the target \ac{AoI} will not be reached at all for the standard case.
Applying our model in this case will result in an even better performance in comparison as the target \ac{AoI} will still be reached (we do not show corresponding results within this manuscript).

The selection of the model parameters as well as the scenario play an important role in the resulting \ac{PAoI} and target \ac{AoI} values.
A configuration which focuses mainly on the distance between vehicles is only useful if the interference domain is already large.
In contrast, if the interference domain is small and observed \ac{AoI} values of neighboring vehicles are similar, a different configuration is useful.
The mobility of vehicles did not play a big role within the results of our simulation study, but it can have a big impact on the knowledge of neighboring vehicles in scenarios with buildings.

Putting our spatial model to practical use within an \ac{AoI}-based beacon adaption algorithm helps reaching an \ac{AoI} requirement while avoiding strong outliers for the beacon rate.
Accordingly, the channel usage is reduced and too infrequent updates are avoided.
It should be noted that we used a very simple approach for controlling the beacon rate based on a \ac{PID} controller.
In fact, a \ac{PID} controller is not well-suited for our use-case of keeping the \ac{PAoI} below the target \ac{AoI} as it tries to reach the set-point (the target \ac{AoI}) exactly and treats positive \& negative deviation from this value equally.
A negative deviation from the set-point (i.e., \ac{PAoI} lower than target \ac{AoI}) is better (and even desired) than a positive deviation (which indicates outdated information).
However, even with our simple approach, we are able to illustrate the benefit of applying the spatial model.

%

\section{Conclusion}%
\label{sec:conclusion}

We explored the use of the \acf{AoI} in the context of \acfp{ITS}.
First approaches of using the \ac{AoI} in \acp{ITS} focused on the original definition only, i.e., to measure the \acp{PAoI} for every received message and then to interpret the resulting values as they are.
We, however, observed that this is not adequate in this application scenario as effects from the wireless communication channel may lead to quite variable \ac{PAoI} measures for farther away vehicles -- even though they play a less important role in many \acp{ITS} applications.

In this paper, we proposed a new way of interpreting the \ac{AoI} for arriving packets.
We focus on the location of the transmitting vehicle as a metric to assess the importance of the information.
Using a weighting coefficient applied to the \ac{PAoI} and also to an \ac{AoI} requirement, we can add a priority measure.
As an example for \acp{ITS}, we use the orientation and the distance of the corresponding vehicles for this process.
It should be noted that the underlying \ac{PAoI} metric is not changed in this procedure, i.e., compatibility with other approach is maintained.
The benefit of applying the model is dependent on its parameters as well as the scenario, which requires careful configuration.
Our spatial model allows to focus on timely updates of relevant vehicles for meeting a given \ac{AoI} requirement, which helps saving resources on the wireless channel.
Applying the model to an \ac{AoI}-based beacon adaption algorithm thus increases information freshness while keeping beacon rates reasonable and resource usage small.

In future work, we plan to apply our spatial model to various different \ac{ITS} applications to gain further insights to the advantages but also limits of the \ac{AoI} metric in general.
Based on the case-study in this work, more sophisticated protocols for adaptively adjusting the beacon rate based on the importance of the information to other vehicles can be build.

%


%

\section{Acknowledgements}

This work was supported in part by
the Federal Ministry of Education and Research (BMBF, Germany) within the 6G Research and Innovation Cluster (6G-RIC) under grant 16KISK020K
and the German Research Foundation (DFG) within the DyMoNet project under grant DR 639/25-1.
The authors would like to thank M. Schettler for the support during the creation of this work.

\bibliographystyle{unsrtnat}
\bibliography{references}

\end{document}